\pdfoutput=1
\documentclass[12pt,a4paper]{article}
\usepackage{ifthen} % for conditional statements
\newboolean{pdflatex}
\setboolean{pdflatex}{true} % False for eps figures 

\newboolean{articletitles}
\setboolean{articletitles}{true} % False removes titles in references

\newboolean{uprightparticles}
\setboolean{uprightparticles}{false} %True for upright particle symbols

\def\paperauthors{LHCb collaboration} % Leave as is for PAPER, CONF and FIGURE
\def\paperasciititle{Search for charge-parity violation in semileptonically tagged D0 to K pi decays} % Set ASCII title here !! MAKE sure it's only ASCII characters !! 
\def\papertitle{Search for charge-parity violation in semileptonically tagged $\Dz\to\Kp\pim$ decays} % Latex formatted title
\def\paperkeywords{{High Energy Physics}, {LHCb}} % Comma separated list
\def\papercopyright{\the\year\ CERN for the benefit of the LHCb collaboration} % new since 9/Apr/2018
\def\paperlicence{CC BY 4.0 licence}
\def\paperlicenceurl{https://creativecommons.org/licenses/by/4.0/}

\newif\ifEnableSectionTOCLinks
\EnableSectionTOCLinksfalse % deactivated

\usepackage[top=1in, bottom=1.25in, left=1in, right=1in]{geometry}

\usepackage{microtype}
\usepackage{lineno}  % for line numbering during review
\usepackage{xspace} % To avoid problems with missing or double spaces after
\usepackage{caption} %these three command get the figure and table captions automatically small

\usepackage{graphicx}  % to include figures (can also use other packages)
\usepackage{color}
\usepackage{colortbl}
\graphicspath{{./figs/}} % Make Latex search fig subdir for figures
\usepackage{amsmath} % Adds a large collection of math symbols
\usepackage{amssymb}
\usepackage{amsfonts}
\usepackage{upgreek} % Adds in support for greek letters in roman typeset

\newcommand*\patchAmsMathEnvironmentForLineno[1]{%
\expandafter\let\csname old#1\expandafter\endcsname\csname #1\endcsname
\expandafter\let\csname oldend#1\expandafter\endcsname\csname
end#1\endcsname
 \renewenvironment{#1}%
   {\linenomath\csname old#1\endcsname}%
   {\csname oldend#1\endcsname\endlinenomath}%
}
\newcommand*\patchBothAmsMathEnvironmentsForLineno[1]{%
  \patchAmsMathEnvironmentForLineno{#1}%
  \patchAmsMathEnvironmentForLineno{#1*}%
}
\AtBeginDocument{%
\patchBothAmsMathEnvironmentsForLineno{equation}%
\patchBothAmsMathEnvironmentsForLineno{align}%
\patchBothAmsMathEnvironmentsForLineno{flalign}%
\patchBothAmsMathEnvironmentsForLineno{alignat}%
\patchBothAmsMathEnvironmentsForLineno{gather}%
\patchBothAmsMathEnvironmentsForLineno{multline}%
\patchBothAmsMathEnvironmentsForLineno{eqnarray}%
}

\usepackage{hyperxmp}

\usepackage[pdftex,
            pdfauthor={\paperauthors},
            pdftitle={\paperasciititle},
            pdfkeywords={\paperkeywords},
            pdfcopyright={Copyright (C) \papercopyright},
            pdflicenseurl={\paperlicenceurl}]{hyperref}
\usepackage[colorinlistoftodos,textsize=scriptsize]{todonotes}

\usepackage[bottom,flushmargin,hang,multiple]{footmisc}

\usepackage[all]{hypcap} % Internal hyperlinks to floats.

\usepackage{xspace} 
\usepackage{upgreek}

\def\lhcb   {\mbox{LHCb}\xspace}

\def\babar  {\mbox{BaBar}\xspace}
\def\belle  {\mbox{Belle}\xspace}

\def\lhc    {\mbox{LHC}\xspace}

\def\MagUp {\mbox{\em Mag\kern -0.05em Up}\xspace}

\ifthenelse{\boolean{uprightparticles}}%
{

 \def\Pmu         {\ensuremath{\upmu}\xspace}

 \def\Ppi         {\ensuremath{\uppi}\xspace}

 \def\PDelta      {\ensuremath{\Delta}\xspace}                 
 \def\PXi         {\ensuremath{\Xi}\xspace}                 
 \def\PLambda     {\ensuremath{\Lambda}\xspace}                 
 \def\PSigma      {\ensuremath{\Sigma}\xspace}                 
 \def\POmega      {\ensuremath{\Omega}\xspace}                 
 \def\PUpsilon    {\ensuremath{\Upsilon}\xspace}
 \let\oldPi\Pi
 \def\PPi         {\ensuremath{\oldPi}\xspace}

 \def\PB      {\ensuremath{\mathrm{B}}\xspace}                 
 \def\PD      {\ensuremath{\mathrm{D}}\xspace}                 

 \def\PK      {\ensuremath{\mathrm{K}}\xspace}                 
 \def\Pb      {\ensuremath{\mathrm{b}}\xspace}                 
 \def\Pc      {\ensuremath{\mathrm{c}}\xspace}                 
 \def\Pd      {\ensuremath{\mathrm{d}}\xspace}                 

 \def\Ps      {\ensuremath{\mathrm{s}}\xspace}                 
                  
 \def\Pu      {\ensuremath{\mathrm{u}}\xspace}                 
 \def\thebaroffset{0.0em}
}
{

 \def\Pmu         {\ensuremath{\mu}\xspace}

 \def\Ppi         {\ensuremath{\pi}\xspace}

 \mathchardef\PDelta="7101
 \mathchardef\PXi="7104
 \mathchardef\PLambda="7103
 \mathchardef\PSigma="7106
 \mathchardef\POmega="710A
 \mathchardef\PUpsilon="7107
 \mathchardef\PPi="7105
 \def\PB      {\ensuremath{B}\xspace}                 
 \def\PD      {\ensuremath{D}\xspace}                 

 \def\PK      {\ensuremath{K}\xspace}                 
 \def\Pb      {\ensuremath{b}\xspace}                 
 \def\Pc      {\ensuremath{c}\xspace}                 
 \def\Pd      {\ensuremath{d}\xspace}                 

 \def\Ps      {\ensuremath{s}\xspace}                 
                  
 \def\Pu      {\ensuremath{u}\xspace}                 
 \def\thebaroffset{0.18em}
}
\newcommand{\offsetoverline}[2][\thebaroffset]{\kern #1\overline{\kern -#1 #2}}%

\makeatletter
\ifcase \@ptsize \relax% 10pt
  \newcommand{\miniscule}{\@setfontsize\miniscule{4}{5}}% \tiny: 5/6
\or% 11pt
  \newcommand{\miniscule}{\@setfontsize\miniscule{5}{6}}% \tiny: 6/7
\or% 12pt
  \newcommand{\miniscule}{\@setfontsize\miniscule{5}{6}}% \tiny: 6/7
\fi
\makeatother

\DeclareRobustCommand{\optbar}[1]{\shortstack{{\miniscule (\rule[.5ex]{1.25em}{.18mm})}
  \\ [-.7ex] $#1$}}

\def\muon       {{\ensuremath{\Pmu}}\xspace}

\def\mun        {{\ensuremath{\Pmu^-}}\xspace} % muon negative (\mum is taken)

\def\uquark    {{\ensuremath{\Pu}}\xspace}
\def\uquarkbar {{\ensuremath{\overline \uquark}}\xspace}

\def\dquark    {{\ensuremath{\Pd}}\xspace}

\def\squark    {{\ensuremath{\Ps}}\xspace}

\def\cquark    {{\ensuremath{\Pc}}\xspace}

\def\bquark    {{\ensuremath{\Pb}}\xspace}

\def\pion   {{\ensuremath{\Ppi}}\xspace}

\def\pip    {{\ensuremath{\pion^+}}\xspace}
\def\pim    {{\ensuremath{\pion^-}}\xspace}
\def\pipm   {{\ensuremath{\pion^\pm}}\xspace}
\def\pimp   {{\ensuremath{\pion^\mp}}\xspace}

\def\kaon    {{\ensuremath{\PK}}\xspace}
\def\KorKbar {\kern \thebaroffset\optbar{\kern -\thebaroffset \PK}{}\xspace}

\def\Kp      {{\ensuremath{\kaon^+}}\xspace}
\def\Km      {{\ensuremath{\kaon^-}}\xspace}
\def\Kpm     {{\ensuremath{\kaon^\pm}}\xspace}
\def\Kmp     {{\ensuremath{\kaon^\mp}}\xspace}
\def\KS      {{\ensuremath{\kaon^0_{\mathrm{S}}}}\xspace}

\def\Dbar    {{\ensuremath{\offsetoverline{\PD}}}\xspace}
\def\D       {{\ensuremath{\PD}}\xspace}

\def\DorDbar {\kern \thebaroffset\optbar{\kern -\thebaroffset \PD}\xspace}
\def\Dz      {{\ensuremath{\D^0}}\xspace}
\def\Dzb     {{\ensuremath{\Dbar{}^0}}\xspace}
\def\Dp      {{\ensuremath{\D^+}}\xspace}
\def\Dm      {{\ensuremath{\D^-}}\xspace}
\def\Dpm     {{\ensuremath{\D^\pm}}\xspace}

\def\DpDm    {\ensuremath{\Dp {\kern -0.16em \Dm}}\xspace}

\def\Dstarp  {{\ensuremath{\D^{*+}}}\xspace}

\def\Dstarpm {{\ensuremath{\D^{*\pm}}}\xspace}

\def\theDstarp{{\ensuremath{\D^{*}(2010)^{+}}}\xspace}

\def\B       {{\ensuremath{\PB}}\xspace}
\def\Bbar    {{\ensuremath{\offsetoverline{\PB}}}\xspace}
\def\Bb      {{\ensuremath{\Bbar}}\xspace}
\def\BorBbar {\kern \thebaroffset\optbar{\kern -\thebaroffset \PB}\xspace}

\def\Bd      {{\ensuremath{\B^0}}\xspace}

\def\BdorBdbar {\kern \thebaroffset\optbar{\kern -\thebaroffset \Bd}\xspace}

\def\Bs      {{\ensuremath{\B^0_\squark}}\xspace}

\def\BsorBsbar {\kern \thebaroffset\optbar{\kern -\thebaroffset \Bs}\xspace}

\def\Y#1S{\ensuremath{\PUpsilon{(#1S)}}\xspace}

\def\LorLbar     {\kern \thebaroffset\optbar{\kern -\thebaroffset \PLambda}\xspace}

\newcommand{\decay}[2]{\mbox{\ensuremath{#1\!\to #2}}\xspace} 

\def\to                 {\ensuremath{\rightarrow}\xspace}

\def\CP                {{\ensuremath{C\!P}}\xspace}

\def\Vud  {{\ensuremath{V_{\uquark\dquark}^{\phantom{\ast}}}}\xspace}

\def\Vus  {{\ensuremath{V_{\uquark\squark}^{\phantom{\ast}}}}\xspace}

\def\Vcds  {{\ensuremath{V_{\cquark\dquark}^\ast}}\xspace}

\def\Vcss  {{\ensuremath{V_{\cquark\squark}^\ast}}\xspace}

\newcommand{\dm}{{\ensuremath{\Delta m}}\xspace}

\def\AT#1     {\ensuremath{A_{\mathrm{T}}^{#1}}\xspace}           % 2

\def\C#1      {\ensuremath{\mathcal{C}_{#1}}\xspace}                       % 9
\def\Cp#1     {\ensuremath{\mathcal{C}_{#1}^{'}}\xspace}                    % 7
\def\Ceff#1   {\ensuremath{\mathcal{C}_{#1}^{\mathrm{(eff)}}}\xspace}        % 9  
\def\Cpeff#1  {\ensuremath{\mathcal{C}_{#1}^{'\mathrm{(eff)}}}\xspace}       % 7
\def\Ope#1    {\ensuremath{\mathcal{O}_{#1}}\xspace}                       % 2
\def\Opep#1   {\ensuremath{\mathcal{O}_{#1}^{'}}\xspace}                    % 7

\newcommand{\ket}[1]{\ensuremath{|#1\rangle}}              % {b}
\newcommand{\nospaceunit}[1]{\ensuremath{\text{#1}}}       
\newcommand{\aunit}[1]{\ensuremath{\text{\,#1}}}       
\newcommand{\tev}{\aunit{Te\kern -0.1em V}\xspace}
\newcommand{\gev}{\aunit{Ge\kern -0.1em V}\xspace}
\newcommand{\mev}{\aunit{Me\kern -0.1em V}\xspace}
\newcommand{\kev}{\aunit{ke\kern -0.1em V}\xspace}
\newcommand{\ev}{\aunit{e\kern -0.1em V}\xspace}
 
\newcommand{\mevc}{\ensuremath{\aunit{Me\kern -0.1em V\!/}c}\xspace}
\newcommand{\gevc}{\ensuremath{\aunit{Ge\kern -0.1em V\!/}c}\xspace}
\newcommand{\mevcc}{\ensuremath{\aunit{Me\kern -0.1em V\!/}c^2}\xspace}
\newcommand{\gevcc}{\ensuremath{\aunit{Ge\kern -0.1em V\!/}c^2}\xspace}
\def\mm   {\aunit{mm}\xspace}

\def\mum  {\ensuremath{\,\upmu\nospaceunit{m}}\xspace}

\def\fb   {\ensuremath{\aunit{fb}}\xspace}
\def\invfb   {\ensuremath{\fb^{-1}}\xspace}

\def\fs   {\aunit{fs}}

\newcommand{\chisq}{\ensuremath{\chi^2}\xspace}
\newcommand{\chisqndf}{\ensuremath{\chi^2/\mathrm{ndf}}\xspace}
\newcommand{\chisqip}{\ensuremath{\chi^2_{\text{IP}}}\xspace}

\def\gsim{{~\raise.15em\hbox{$>$}\kern-.85em
          \lower.35em\hbox{$\sim$}~}\xspace}
\def\lsim{{~\raise.15em\hbox{$<$}\kern-.85em
          \lower.35em\hbox{$\sim$}~}\xspace}

\def\sPlot{\mbox{\em sPlot}\xspace}

\def\pt         {\ensuremath{p_{\mathrm{T}}}\xspace}

\def\ptot       {\ensuremath{p}\xspace}

\def\degrees{\ensuremath{^{\circ}}\xspace}

\def\evtgen     {\mbox{\textsc{EvtGen}}\xspace}

\def\geant      {\mbox{\textsc{Geant4}}\xspace}

\def\photos     {\mbox{\textsc{Photos}}\xspace}

\def\pythia     {\mbox{\textsc{Pythia}}\xspace}

\def\tell1  {TELL1\xspace}
\def\ukl1   {UKL1\xspace}

\newcommand{\lhcborcid}[1]{\href{https://orcid.org/#1}{\hspace*{0.1em}\raisebox{-0.45ex}{\includegraphics[width=1em]{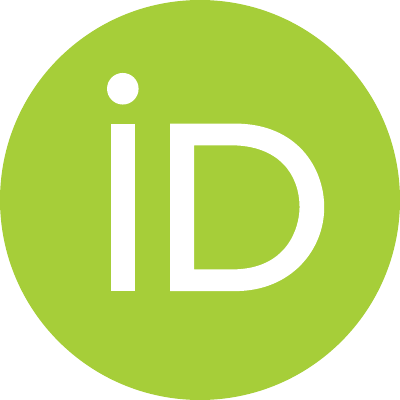}}}}

\hypersetup{
  colorlinks   = true, %Colours links instead of ugly boxes
  urlcolor     = blue, %Colour for external hyperlinks
  linkcolor    = blue, %Colour of internal links
  citecolor    = red   %Colour of citations
}

\ifEnableSectionTOCLinks
    \usepackage[explicit]{titlesec} % to change headings
    
    \let\oldcontentsline\contentsline
    \renewcommand

    \titleformat{\section}{\normalfont\Large\bf}{\hyperlink{tocsection.\thesection}{{\thesection} \parbox[t]{\dimexpr\textwidth-1pc}{#1}}}{1pc}{}

    \titleformat{\subsection}{\normalfont\bf}{\hyperlink{tocsubsection.\thesubsection}{{\thesubsection} \parbox[t]{\dimexpr\textwidth-1pc}{#1}}}{1pc}{}

    \titleformat{name=\section,numberless}[display]{}{}{0pt}{\normalfont\Huge\bfseries #1}
\fi

\usepackage{cite} % Allows for ranges in citations
\usepackage{mciteplus}
\def\spion   {{\ensuremath{\pion_{\mathrm{s}}^{}}}\xspace}

\usepackage{subcaption}
\usepackage{multirow}
\usepackage{makecell}
\usepackage{siunitx}
\usepackage{array}

\usepackage[normalem]{ulem}

\newcolumntype{M}{>{\raggedleft\arraybackslash}m{3.7cm}}
\newcolumntype{R}{>{\raggedleft\arraybackslash}m{5cm}}

\def\spion   {{\ensuremath{\pion_{\mathrm{s}}^{}}}\xspace}

\def\RDp   {{\ensuremath{R_{D}^{+}}}\xspace}
\def\RDm   {{\ensuremath{R_{D}^{-}}}\xspace}
\def\RDpm   {{\ensuremath{R_{D}^{\pm}}}\xspace}
\def\yp   {{\ensuremath{y^{\prime+}}}\xspace}
\def\ym   {{\ensuremath{y^{\prime-}}}\xspace}
\def\ypm   {{\ensuremath{y^{\prime\pm}}}\xspace}
\def\y   {{\ensuremath{y^{\prime}}}\xspace}

\def\xpm   {{\ensuremath{x^{\prime\pm}}}\xspace}

\def\xsqp   {{\ensuremath{(x^{\prime+})^{2}}}\xspace}
\def\xsqm   {{\ensuremath{(x^{\prime-})^{2}}}\xspace}
\def\xsqpm   {{\ensuremath{(x^{\prime\pm})^{2}}}\xspace}
\def\xsq   {{\ensuremath{(x^{\prime})^{2}}}\xspace}

\def\RD   {{\ensuremath{R_{D}}}\xspace}
\def\ckp   {{\ensuremath{c_{\kaon\pion}^{\phantom{\prime}}}}\xspace}
\def\ckpp   {{\ensuremath{c_{\kaon\pion}^{\prime}}}\xspace}
\def\AD   {{\ensuremath{A_{D}}}\xspace}

\def\Dckp   {{\ensuremath{\Delta{c_{\kaon\pion}^{\phantom{\prime}}}}}\xspace}

\def\Dckpp   {{\ensuremath{\Delta{c_{\kaon\pion}^{\prime}}}}\xspace}

\usepackage{longtable} % only for template; not usually to be used in PAPERs

\begin{document}

%%%%%%%%%%%%%%%%%%%%%%%%%
%%%%% Title     %%%%%%%%%
%%%%%%%%%%%%%%%%%%%%%%%%%
\renewcommand{\thefootnote}{\fnsymbol{footnote}}
\setcounter{footnote}{1}

% %%%%%%% CHOOSE TITLE PAGE--------
%\onecolumn
%\input{title-LHCb-INT}
%\input{title-LHCb-ANA}
%\input{title-LHCb-CONF}
%\input{title-LHCb-FIGURE}
% ===============================================================================
% Purpose: LHCb-PAPER journal paper title page template
% Author: 
% Created on: 2010-09-25
% ===============================================================================

%%%%%%%%%%%%%%%%%%%%%%%%%
%%%%%  TITLE PAGE  %%%%%%
%%%%%%%%%%%%%%%%%%%%%%%%%
\begin{titlepage}
\pagenumbering{roman}

% Header ---------------------------------------------------
\vspace*{-1.5cm}
\centerline{\large EUROPEAN ORGANIZATION FOR NUCLEAR RESEARCH (CERN)}
\vspace*{1.5cm}
\noindent
\begin{tabular*}{\linewidth}{lc@{\extracolsep{\fill}}r@{\extracolsep{0pt}}}
\ifthenelse{\boolean{pdflatex}}% Logo format choice
{\vspace*{-1.5cm}\mbox{\!\!\!\includegraphics[width=.14\textwidth]{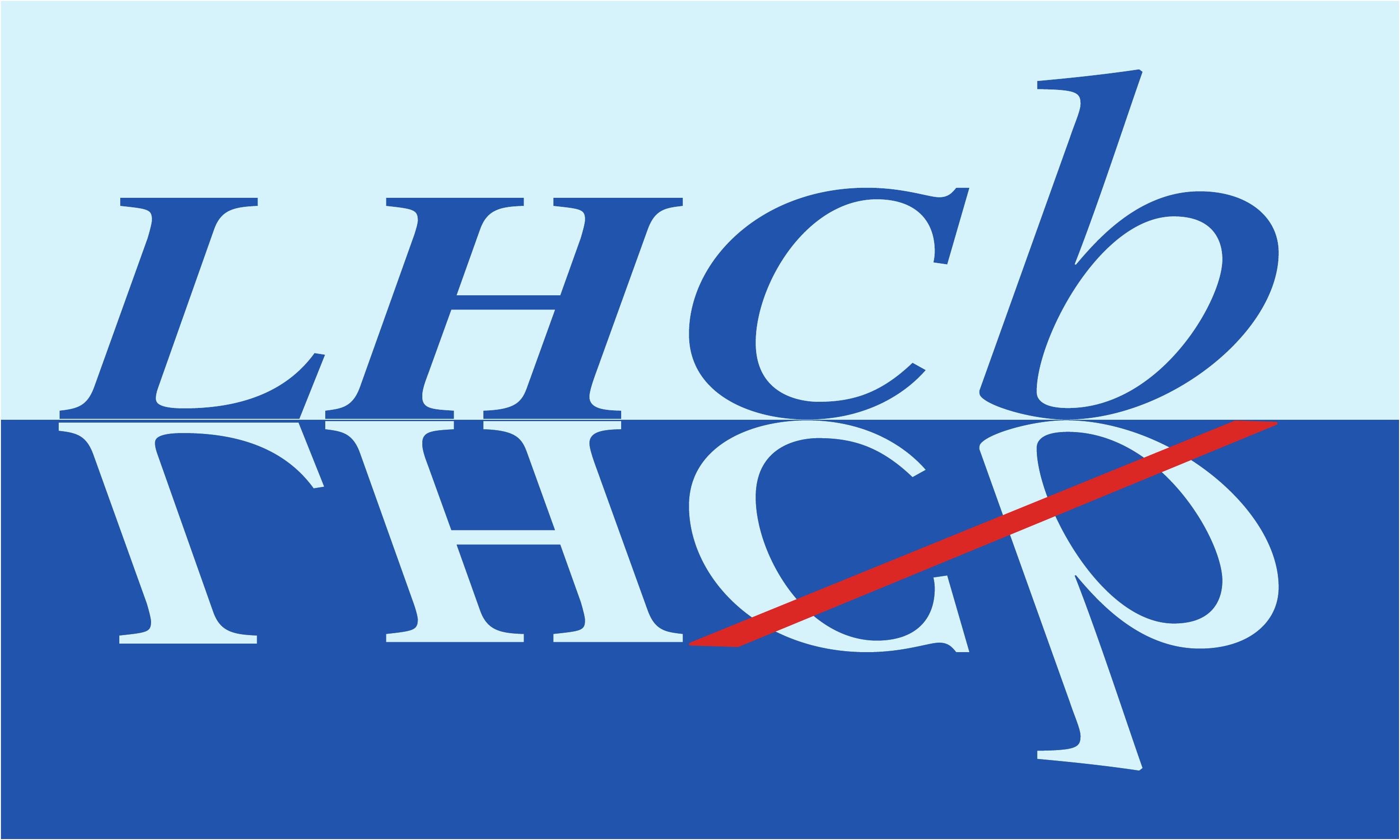}} & &}%
{\vspace*{-1.2cm}\mbox{\!\!\!\includegraphics[width=.12\textwidth]{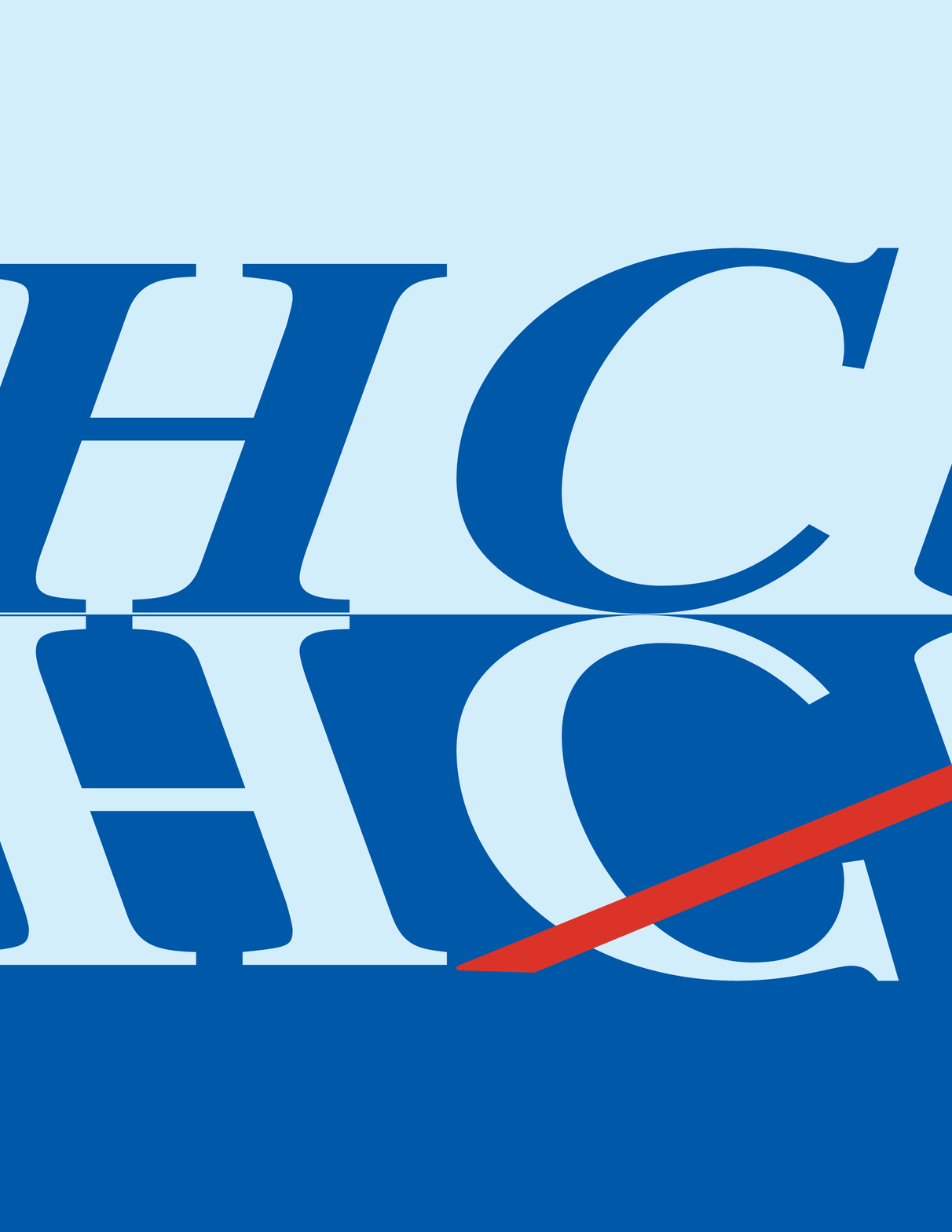}} & &}%
\\
 & & CERN-EP-2024-319 \\  % ID 
 & & LHCb-PAPER-2024-044 \\  % ID 
 & & January 20, 2025 \\ %\today \\ % Date - Can also hardwire e.g.: 23 March 2010
 & & \\
% not in paper \hline
\end{tabular*}

\vspace*{4.0cm}

% Title --------------------------------------------------
{\normalfont\bfseries\boldmath\huge
\begin{center}
% DO NOT EDIT HERE. Instead edit macro in main.tex to keep metadata correct
  \papertitle 
\end{center}
}

\vspace*{1.8cm}

% Authors -------------------------------------------------
\begin{center}
%In the footnote, replace 'paper' by 'Letter' in case of submission to PRL or PLB 
% Edit macro in main.tex to keep metadata correct
\paperauthors\footnote{Authors are listed at the end of this paper.}
\end{center}

\vspace{\fill}

% Abstract -----------------------------------------------
\begin{abstract}
  \noindent
An analysis of the flavour oscillations of the charmed neutral meson is presented. The ratio of ${\Dz\to\Kp\pim}$ and ${\Dz\to\Km\pip}$ decay rates is measured as a function of the decay time of the $\Dz$ meson and compared with the charge-conjugated system to search for charge-parity violation. The meson flavour at production is double-tagged by the charges of the muon and pion in the preceding ${\ensuremath{\kern 0.18em \overline{\kern -0.18em \PB}}\to{\D^{*}(2010)^{+}}\mun X}$ and ${{\D^{*}(2010)^{+}}\to\Dz\pip}$ decays, respectively. These decays are selected from proton-proton collision data collected by the LHCb experiment at a centre-of-mass energy of ${13\tev}$ and corresponding to an integrated luminosity of ${5.4\invfb}$. The flavour oscillation parameters, relating to the differences in mass and width of the mass eigenstates, are found to be ${y^\prime=(5.8\pm1.6)\times10^{-3}}$ and ${(x^\prime)^2=(0.0\pm1.2)\times10^{-4}}$.
No evidence for charge-parity violation is seen either in the flavour oscillations or in the decay, where the direct charge-parity asymmetry is measured to be ${A_{D}=(2.3\pm1.7)\,{\%}}$.
\end{abstract}

\vspace*{1.8cm}

\begin{center}
  Published in
  JHEP 03 (2025) 149
\end{center}

\vspace{\fill}

{\footnotesize 
% Edit macro in main.tex to keep metadata correct
\centerline{\copyright~\papercopyright. \href{\paperlicenceurl}{\paperlicence}.}}
\vspace*{2mm}

\end{titlepage}

%%%%%%%%%%%%%%%%%%%%%%%%%%%%%%%%
%%%%%  EOD OF TITLE PAGE  %%%%%%
%%%%%%%%%%%%%%%%%%%%%%%%%%%%%%%%

%  empty page follows the title page ----
\newpage
\setcounter{page}{2}
\mbox{~}
%\newpage
%
%% Author List ----------------------------
%%  You need to get a new author list!
%\input{LHCb_authorlist.tex}
%
%The author list for journal publications is provided by the Membership Committee shortly after 'approval to go to paper' has been given.
%%It will be made available on the page
%%\verb!http://www.physik.uzh.ch/~strauman/forMemCo/LHCb-PAPER-XXXX-XXX/! .
%It will be sent to you by email shortly after a paper number has beens assigned.
%The author list should be included already at first circulation, 
%to allow new members of the collaboration to verify whether they have been included correctly.
%Occasionally a misspelled name is corrected or associated institutions become full members.
%In that case, a new author list will be sent to you.
%In case line numbering doesn't work well after including the authorlist, try moving the \verb!\bigskip! after the last author to a separate line.
%
%
%The authorship for Conference Reports should be ``The LHCb
%  collaboration'', with a footnote giving the name(s) of the contact
%  author(s), but without the full list of collaboration names.

%\twocolumn
% %%%%%%%%%%%%% ---------

\renewcommand{\thefootnote}{\arabic{footnote}}
\setcounter{footnote}{0}

%%%%%%%%%%%%%%%%%%%%%%%%%%%%%%%%
%%%%%  Table of Content   %%%%%%
%%%%%%%%%%%%%%%%%%%%%%%%%%%%%%%%
%%%% Uncomment if desired
%\tableofcontents

\cleardoublepage

%%%%%%%%%%%%%%%%%%%%%%%%%
%%%%% Main text %%%%%%%%%
%%%%%%%%%%%%%%%%%%%%%%%%%

\pagestyle{plain} % restore page numbers for the main text
\setcounter{page}{1}
\pagenumbering{arabic}

%% Uncomment during review phase. 
%% Comment before a final submission.
% \linenumbers

%% This is the main body
%% It is useful to have a single file so comments are not missed in overleaf.
\section{Introduction}
\label{sec:introduction}

In the Standard Model (SM) violation of the charge-parity (\CP) symmetry is theorised in both the weak and strong sectors, but is only experimentally observed in the weak interactions of quarks.
Neutral-meson decays are attractive systems for \CP-violation studies since neutral final states are accessible to both flavour eigenstates.
Mixing, the ability of neutral mesons to oscillate between the flavour eigenstates due to the nonzero mass and width differences of the mass eigenstates, provides an additional path for \CP violation to manifest.
In the SM mixing is possible via flavour-changing neutral currents (FCNCs) but is suppressed by the Glashow--Iliopoulous--Maiani mechanism~\cite{GIMMechanism} and the corresponding Cabibbo--Kobayashi--Maskawa (CKM) matrix elements~\cite{NeutralMesonMixings}.
This is especially the case in charm hadrons since the suppression is not compensated by a heavy quark mass in the FCNC, as it is in strange and beauty systems.

Charmed-hadron decays to two oppositely charged hadrons of different type can be studied for both mixing and \CP violation. There are three types of \CP violation that can occur: in the decay, in the mixing and in the interference between mixing and decay. The weak decay of the \Dz meson\footnote{The same is true for the charge-conjugate system hereafter unless stated otherwise.}
(quark content ${\cquark\uquarkbar}$) to ${\Km\pip}$ mesons is dominated by a Cabibbo-favoured (CF) tree amplitude, ${\mathcal{A}_{\Km\pip}}$, proportional to the product of CKM matrix elements ${\Vcss\Vud}$.
The decay to the charge-conjugate ${\Kp\pim}$ state is much less frequent with two main contributions of comparable amplitudes: the double-Cabibbo-suppressed (DCS) direct decay amplitude, ${\mathcal{A}_{\Kp\pim}}$, proportional to ${\Vcds\Vus}$, and the CF decay of the \Dzb meson, ${\overline{\mathcal{A}}_{\Kp\pim}}$, following a (suppressed) flavour oscillation from \Dz to \Dzb meson. The \Dz decays to ${\Km\pip}$ and ${\Kp\pim}$ are named right-sign (RS) and wrong-sign (WS), respectively, due to their relative rates.
    
The time-dependent decay rates are found from ${\left|\langle{\Kmp\pipm}|\mathbf{H}|\Dz(t)\rangle\right|^{2}}$, where ${\Dz(t)}$ denotes the decay-time ${t}$ evolution of a meson with \Dz flavour at production and ${\mathbf{H}}$ is the effective Hamiltonian of the flavour oscillations. Expanded to second order in ${t}$, the decay rates of the neutral mesons become~\cite{aburke}
\begin{align}
    \left|\langle{\Km\pip}|\mathbf{H}|\Dz(t)\rangle\right|^{2} &= \mathcal{N} e^{-t/\tau_{\Dz}}\left|\mathcal{A}_{\Km\pip}\right|^{2}, \\
    \left|\langle{\Kp\pim}|\mathbf{H}|\Dz(t)\rangle\right|^{2} &= \mathcal{N} e^{-t/\tau_{\Dz}}\left|\overline{\mathcal{A}}_{\Kp\pim}\right|^{2} R^{+}(t), \\
    \left|\langle{\Kp\pim}|\mathbf{H}|\Dzb(t)\rangle\right|^{2} &= \mathcal{N} e^{-t/\tau_{\Dz}}\left|\overline{\mathcal{A}}_{\Kp\pim}\right|^{2}, \\
    \left|\langle{\Km\pip}|\mathbf{H}|\Dzb(t)\rangle\right|^{2} &= \mathcal{N} e^{-t/\tau_{\Dz}}\left|\mathcal{A}_{\Km\pip}\right|^{2} R^{-}(t),
\end{align}
where ${\tau_{\Dz}=410.3\fs}$~\cite{PDG2024} is the \Dz mean lifetime, and $\mathcal{N}$ is a normalisation factor. The ${R^\pm(t)}$ terms in the WS decay rates are known as the quadratic mixing approximation and are written as
\begin{equation}
    R^{\pm}(t) = \RDpm + \sqrt{\RDpm}\ypm\left(\frac{t}{\tau_{\Dz}}\right) + \frac{\xsqpm+(\ypm)^{2}}{4}\left(\frac{t}{\tau_{\Dz}}\right)^{2},
    \label{eqn: quadratic mixing approximation}
\end{equation}
where
\begin{align}
    \RDp &\equiv \left|\frac{\mathcal{A}_{\Kp\pim}}{\overline{\mathcal{A}}_{\Kp\pim}}\right|^{2}, \;\; \RDm \equiv \left|\frac{\overline{\mathcal{A}}_{\Km\pip}}{\mathcal{A}_{\Km\pip}}\right|^{2}, \\
    \begin{pmatrix}
        \xpm \\
        \ypm
    \end{pmatrix}
    &=\left|q/p\right|^{\pm1}
    \begin{pmatrix}
        \cos{[\delta\pm\phi]} & \sin{[\delta\pm\phi]} \\
        -\sin{[\delta\pm\phi]} & \cos{[\delta\pm\phi]}
    \end{pmatrix}
    \begin{pmatrix}
        x \\
        y 
    \end{pmatrix}.
    \label{eqn:x'y'+-}
\end{align}
Here, the mixing parameters $x$ and $y$ are proportional to the differences between the mass eigenstates $D_{1,2}$ in mass ${\Delta{M}\equiv M_2-M_1}$ and width ${\Delta{\Gamma}\equiv \Gamma_2-\Gamma_1}$, respectively,
\begin{align}
    x &\equiv \frac{\Delta{M}}{\overline{\Gamma}} \label{eqn:x}, \\
    y &\equiv \frac{\Delta{\Gamma}}{2\overline{\Gamma}},
\end{align}
where the average decay width $\overline{\Gamma}$ is given by
\begin{equation}
  \overline{\Gamma} \equiv \frac{\Gamma_{1} + \Gamma_{2}}{2} \label{eqn:avg-lifetime}.
\end{equation}
The angle $\delta$ is the \CP-conserving strong-phase difference between the DCS and CF decays, ${\delta\equiv \arg(\overline{\mathcal{A}}_{\Km\pip}/\mathcal{A}_{\Km\pip})}$, and ${\phi\equiv \arg\left(q/p\right)}$ is the \CP-violating weak-phase difference between the DCS and CF decays, where $q$ and $p$ are complex coefficients that define the mass eigenstates
\begin{equation}
    |D_{1,2}\rangle = p|\Dz\rangle \pm q|\Dzb\rangle,
\end{equation}
where the subscript $1$ refers to the $+$ sign and the phase convention is chosen such that $\CP \ket{\Dz} = - \ket{\Dzb}$ and, in the case of \CP\ symmetry, $D_2$ is \CP\ even.

Violation of \CP symmetry manifests in the decay when ${|\mathcal{A}_{f}|\neq|\overline{\mathcal{A}}_{\bar{f}}|}$ and in mixing when ${\left|q/p\right|\neq1}$. Assuming \CP symmetry in RS decays only (${|\mathcal{A}_{\Km\pip}|=|\overline{\mathcal{A}}_{\Kp\pim}|}$), the time-dependent ratios of WS-to-RS decay rates give access to the mixing parameters of each flavour,
\begin{align}
    \frac{\left|\langle{\Kp\pim}|\mathbf{H}|\Dz(t)\rangle\right|^{2}}{\left|\langle{\Km\pip}|\mathbf{H}|\Dz(t)\rangle\right|^{2}} &= R^{+}(t), \label{eqn: R+} \\
    \frac{\left|\langle{\Km\pip}|\mathbf{H}|\Dzb(t)\rangle\right|^{2}}{\left|\langle{\Kp\pim}|\mathbf{H}|\Dzb(t)\rangle\right|^{2}} &= R^{-}(t). \label{eqn: R-}
\end{align}
The \Dz\ mixing parameters are very small ($<1\aunit{\%}$) in the SM, with evidence of mixing not published until 2007 by \babar and \belle~\cite{BABARMixing,BelleMixing}. The no-mixing hypothesis is now excluded by a significance exceeding nine standard deviations by \lhcb using ${\Dz\to\Kmp\pipm}$ decays alone~\cite{LHCb-PAPER-2016-033,LHCb-PAPER-2024-008}, with further individual evidence for nonzero $x$ and $y$ values published using ${\Dz\to\KS\pip\pim}$ decays~\cite{LHCb-PAPER-2019-001,LHCb-PAPER-2021-009,LHCb-PAPER-2022-020} and \decay{\Dz}{h^+h^-} ($h=\kaon,\pion$) decays~\cite{LHCb-PAPER-2018-038,LHCb-PAPER-2021-041}, respectively.

In this measurement we analyse semileptonic decays of $\Bb$ hadrons, ${\Bb\to\theDstarp\mun{X}}$, where $X$ represents unreconstructed particles.\footnote{Hereafter, the \theDstarp meson is denoted as \Dstarp.}
The charges of the muon and the spectator pion in the subsequent ${\Dstarp\to\Dz\pip}$ decay enable the flavour of the $\Dz$ meson to be tagged at production time, $t=0$, and are, therefore, necessary to classify its decay to ${\Kmp\pipm}$ final state as either right- or wrong-sign.
The spectator pion has a low energy since the available phase space in the ${\Dstarp\to\Dz\pip}$ decay is limited.
As such the pion is typically named ``soft'' and given the subscript $s$.
The full decay chain therefore has two oppositely charged tagging particles and is referred to as double-tagged.
Double- and \Dstarp-tagged samples are also referred to as secondary and prompt, respectively, in relation to the time elapsed between the \lhc proton-proton ($pp$) interaction and the production of the $\Dz$ meson.
Secondary decays are delayed by the non-negligible lifetime of \bquark\ hadrons.
While much smaller in sample size, the benefit of studying double-tagged decays is a larger purity and a gain in sensitivity in the low $\Dz$ decay-time region; this region is not accessible in prompt decays since the \Dz decay vertex is required not to coincide with the highly congested $pp$ interaction region.

We present a measurement of the $\Dz$ mixing and \CP-violating parameters using double-tagged decays collected by the \lhcb experiment during Run~2 operation of the \lhc.
The dataset is obtained from ${5.4\invfb}$ of integrated luminosity following $pp$ collisions at a centre-of-mass energy of ${13\tev}$ recorded between 2016 and 2018.
The dataset is analysed under three different measurement hypotheses.
The baseline measurement allows all forms of \CP violation, with individual sets of parameters for $\Dz$ and $\Dzb$ decays.
We then constrain the decay rates, $\RDp$ and $\RDm$, to be shared between subsamples, enforcing \CP\ symmetry in the decay.
Finally, we apply further constraints to measure mixing only, imposing \CP conservation in both the decay and mixing.
In order to avoid experimenter's bias, the results of the analysis were not examined until the full procedure had been finalised.
The measurement strategy closely follows that of Ref.~\cite{LHCb-PAPER-2016-033}, which corresponds to a smaller dataset collected during Run~1.
A combination of \lhcb Run~1 and Run~2 results for double-tagged ${\Dz\to\Kmp\pipm}$ decays is also performed.
Analogous measurements with prompt decays are published in Ref.\cite{LHCb-PAPER-2024-008} for Run~1 and Run~2.

\section{\lhcb detector}
\label{sec:Detector}

The \lhcb detector~\cite{LHCb-DP-2008-001,LHCb-DP-2014-002} is a single-arm forward
spectrometer covering the \mbox{pseudorapidity} range ${2<\eta <5}$,
designed for the study of particles containing \bquark or \cquark
quarks. The detector includes a high-precision tracking system
consisting of a silicon-strip vertex detector surrounding the $pp$
interaction region~\cite{LHCb-DP-2014-001}, a large-area silicon-strip detector located
upstream of a dipole magnet with a bending power of about
${4{\mathrm{\,T\,m}}}$, and three stations of silicon-strip detectors and straw
drift tubes~\cite{LHCb-DP-2017-001}
placed downstream of the magnet.
The tracking system provides a measurement of the momentum, \ptot, of charged particles with
a relative uncertainty that varies from 0.5\,\% at low momentum to 1.0\,\% at ${200\gevc}$.
The minimum distance of a track to a primary $pp$ collision vertex (PV), the impact parameter (IP), 
is measured with a resolution of ${(15+29/\pt)\mum}$,
where \pt is the component of the momentum transverse to the beam, in\,\gevc.
Different types of charged hadrons are distinguished using information
from two ring-imaging Cherenkov detectors (RICH)~\cite{LHCb-DP-2012-003}. 
Photons, electrons and hadrons are identified by a calorimeter system consisting of
scintillating-pad and preshower detectors, an electromagnetic
and a hadronic calorimeter. Muons are identified by a
system composed of alternating layers of iron and multiwire
proportional chambers~\cite{LHCb-DP-2012-002}.

The online event selection is performed by a trigger which consists of a hardware stage followed by a two-level software stage.
In between the two software stages, an alignment and calibration of the detector is performed in near real-time and their results are used in the trigger~\cite{LHCb-PROC-2015-011}.
The same alignment and calibration information is propagated to the offline reconstruction, ensuring consistent and high-quality particle identification (PID) information between the trigger and offline software.
The identical performance of the online and offline reconstruction offers the opportunity to perform physics analyses directly using candidates reconstructed in the trigger \cite{LHCb-DP-2019-001,LHCb-DP-2019-002} which the present analysis exploits.
The storage of only the triggered candidates enables a reduction in the event size by an order 
of magnitude.

Simulation is used to model both the background components and the signal decay-time resolution.
In the simulation, $pp$ collisions are generated using
\pythia~\cite{Sjostrand:2007gs,*Sjostrand:2006za} 
with a specific \lhcb configuration~\cite{LHCb-PROC-2010-056}.
Decays of unstable particles
are described by \evtgen~\cite{Lange:2001uf}, in which final-state
radiation is generated using \photos~\cite{davidson2015photos}.
The interaction of the generated particles with the detector, and its response,
are implemented using the \geant
toolkit~\cite{Allison:2006ve, *Agostinelli:2002hh} as described in
Ref.~\cite{LHCb-PROC-2011-006}. 
The underlying $pp$ interaction is reused multiple times, each with an independently generated signal decay~\cite{LHCb-DP-2018-004}.

\section{Candidate selection}
\label{sec:selection}

Candidate \Dz mesons are first formed by combining two opposite-charge kaon and pion tracks before consecutively adding soft pion and muon tagging tracks to form \Dstarp and \Bb candidates, respectively.
In the offline selection, trigger signals are associated with reconstructed particles. Selection requirements can therefore be made on the trigger selection itself and on whether the decision was due to the signal candidate, other particles produced in the $pp$ collision, or a combination of both. 
The hardware trigger selects candidates with a high-\pt muon track through the downstream muon tracking stations.
The muon candidates must then pass further \pt and \chisqip criteria in the first-stage software trigger, where \chisqip\ is defined as the difference in the vertex-fit \chisq of a given PV, reconstructed with and without the track under consideration.
Using RICH information, the tracks which form the \Dz candidate are assigned PID hypotheses which are required to be consistent with those of a kaon and a pion by strict criteria.
Similar track PID requirements are imposed on the soft pion and muon tracks that subsequently form the \Dstarp and \Bb candidates.
During the second-stage software trigger, high-momentum selections are applied to all particle tracks. The kaon and pion are required to satisfy ${p>2\gevc}$ and ${\pt>200\mevc}$.
The soft pion momentum requirement is the same but has a slightly looser transverse momentum selection of $\pt>150\mev$, while muons have a stricter ${p>3\gevc}$ and ${\pt>1\gevc}$ acceptance.
Kaon, pion and muon tracks are required to satisfy ${\chisqip>9}$ with respect to the associated PV\footnote{The PV that fits best to the flight direction of the \Bb candidate is taken as the associated PV\@.}.
Windows on the invariant mass of each parent candidate are applied around the known values~\cite{PDG2024}, and a maximum vertex \chisq per degree of freedom, \chisqndf, is set.
Lastly, the momentum of the \Bb candidate must align, within 2.6\degrees, with the line joining its decay vertex to the PV for the candidate to pass the trigger selection.

Further selections are applied offline to the candidates which pass trigger requirements.
The PID requirements on the \kaon and \pion tracks are tightened considerably to minimise random reconstruction of \Dz candidates from misidentified tracks.
Candidates with misidentified final-state particles are suppressed by requiring that the invariant mass $m(\kaon\pion)$ is within three times the resolution, $\sigma_{m(\kaon\pion)}=8\mevcc$, of the known value, $m_\Dz$~\cite{PDG2024}.
In addition, the invariant masses calculated under the hypotheses that one particle has been misidentified, $m(\kaon\kaon)$ and $m(\pion\pion)$, must be outside a window of $5\sigma_{m(\kaon\pion)}$ centred on $m_\Dz$.
Influence from material interactions on the mixing dynamics is reduced by vetoing candidates whose \Dz decay vertex lies more than 4.5\mm from the PV.
A multivariate classifier~\cite{Breiman} is used to determine the probability that the \spion track is a reconstruction of random hits in the tracking subdetectors on which a maximum probability of 0.35 is set.
This is paired with a further restriction on the probability of the soft pion being identified as an electron.
A similar multivariate classifier, which determines the probability of muon PID, is applied to the muon track with a minimum probability set at 0.4.
The \chisqndf of the global fit to all trajectories and vertices in the \Bb decay chain is required to be less than 100.
Finally, a mass window ${3100<m(\Bb)<5100\mevcc}$ is set, where $m(\Bb)$ is the invariant mass of the \Bb candidate calculated from its detectable decay products.

The magnetic field deflects oppositely charged particles in opposite directions and this can lead to detection asymmetries.
Periodically reversing the magnetic field polarity throughout the data taking almost cancels the effect.
The configuration with the magnetic field pointing upwards (downwards) bends positively (negatively) charged particles in the horizontal plane towards the centre of the LHC ring.
We study instrumental \spion and \muon detection asymmetries that are residual to the cancellation obtained by reversing the magnetic field polarity.
The regions of the phase space of the components of momenta in the directions transverse and longitudinal to the beam featuring a large (${>99\,\%}$) asymmetry are removed to avoid significant biases to the direct \CP asymmetry~\cite{LHCb-PAPER-2024-008}.
A dedicated study of the \kaon-\pion detection efficiency asymmetry, $A_{\kaon\pion}$, is discussed in Sec.~\ref{sec: asymmetry}.

A multivariate classifier~\cite{Breiman} is trained to maximise the significance, $S=N_{S}/\sqrt{N_{S}+N_{B}}$, where $N_{S(B)}$ is the signal (background) yield, in the WS sample.
The variables chosen to train the classifier are determined from a data-driven approach.
These are the $x$ and $y$ coordinates of the \Dz decay vertex, its IP with respect to the \Dstarp decay vertex and the flight distance and lifetime \chisq values of the \Dstarp meson.
The sample labelled as signal in the training stage consists of both WS and RS candidates which occupy the signal region of the distribution of the difference between the reconstructed \Dstarp and \Dz invariant masses, \dm.
The sample labelled as background consists of candidates that have been tagged by a muon and a soft pion with the same charge which also lie outside of the signal regions of both the \dm and $m(\kaon\pion)$ distributions.
This background contains mainly combinations of random tracks, therefore referred to as combinatorial background, resulting in smoothly varying distributions.
To avoid losing data to training, the datasets are $k$-folded, resulting in multiple trained classifiers with around $70\aunit{\%}$ background-rejection efficiency and $80(30)\aunit{\%}$ RS (WS) signal-selection efficiency from each.

The decay time of the \Dz meson is calculated as $t=m_\Dz d / p$, where $p$ is its momentum and $d$ its flight distance, determined from the \Dz production and decay vertices.
The measured \Dz decay-time distribution is smeared into the negative decay time region by the resolution of the \lhcb detector.
The ${t<-0.5\tau_{\Dz}}$ region is dominated by prompt and combinatorial background contributions and is therefore removed from the sample.

Events may contain more than one candidate through a number of mechanisms, for example, from the association of numerous soft pion tracks to form multiple \Dstarpm candidates from a single \Dz candidate, where at most one can be genuine signal.
Incorrect combinations of this nature can populate both the RS and WS samples with the same \Dz candidate and ultimately bias the mixing parameters.
In events where multiple candidates are found, one is retained at random and the others discarded.
This reduces the entire sample by $1.5\,\%$.
Less than $0.02\,\%$ of candidates in these events appear in both the WS and RS samples.

\begin{figure}
    \centering
     \begin{subfigure}[b]{0.45\textwidth}
         \centering
         \includegraphics[width=\textwidth]{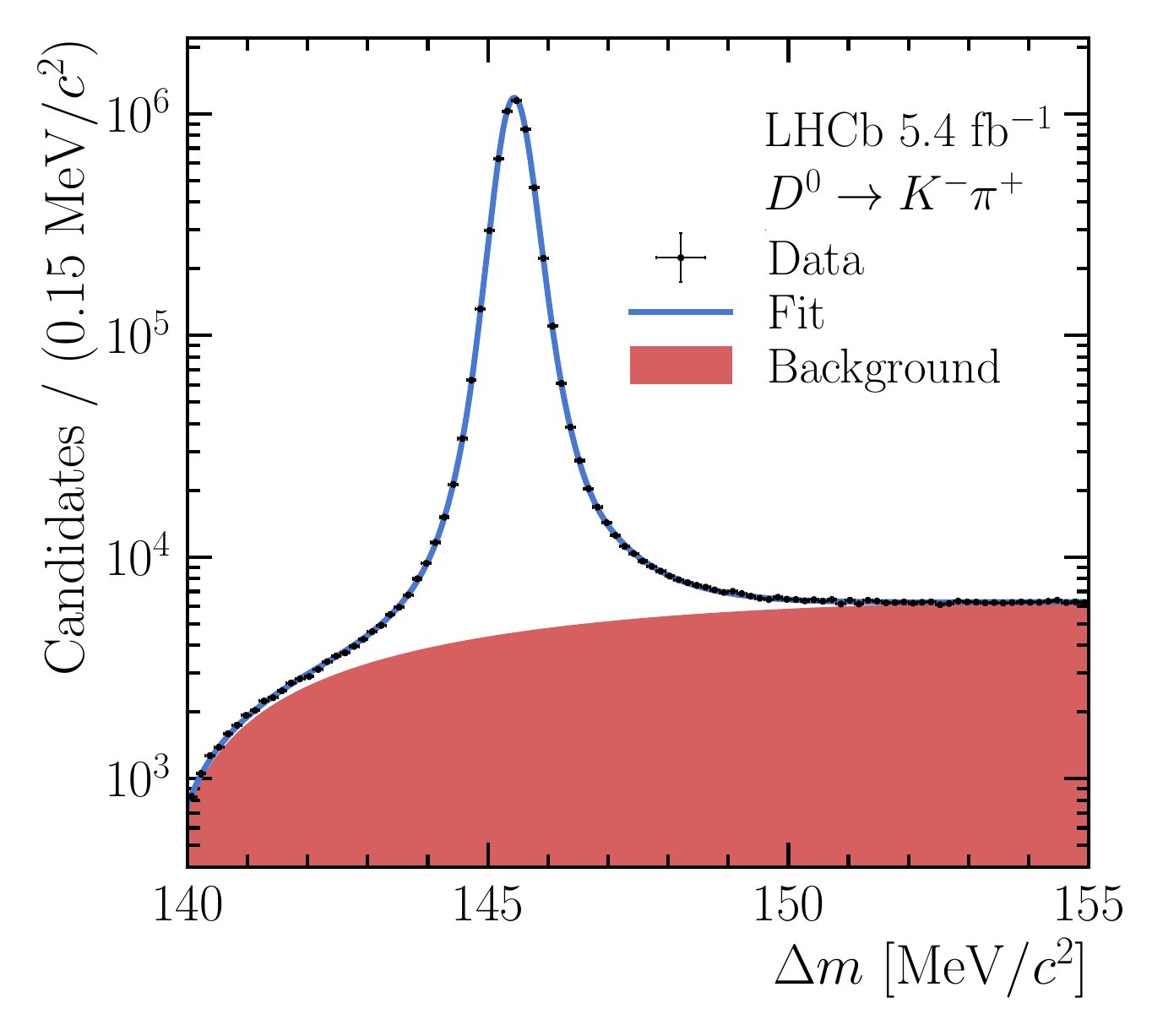}
     \end{subfigure}
     \begin{subfigure}[b]{0.45\textwidth}
         \centering
         \includegraphics[width=\textwidth]{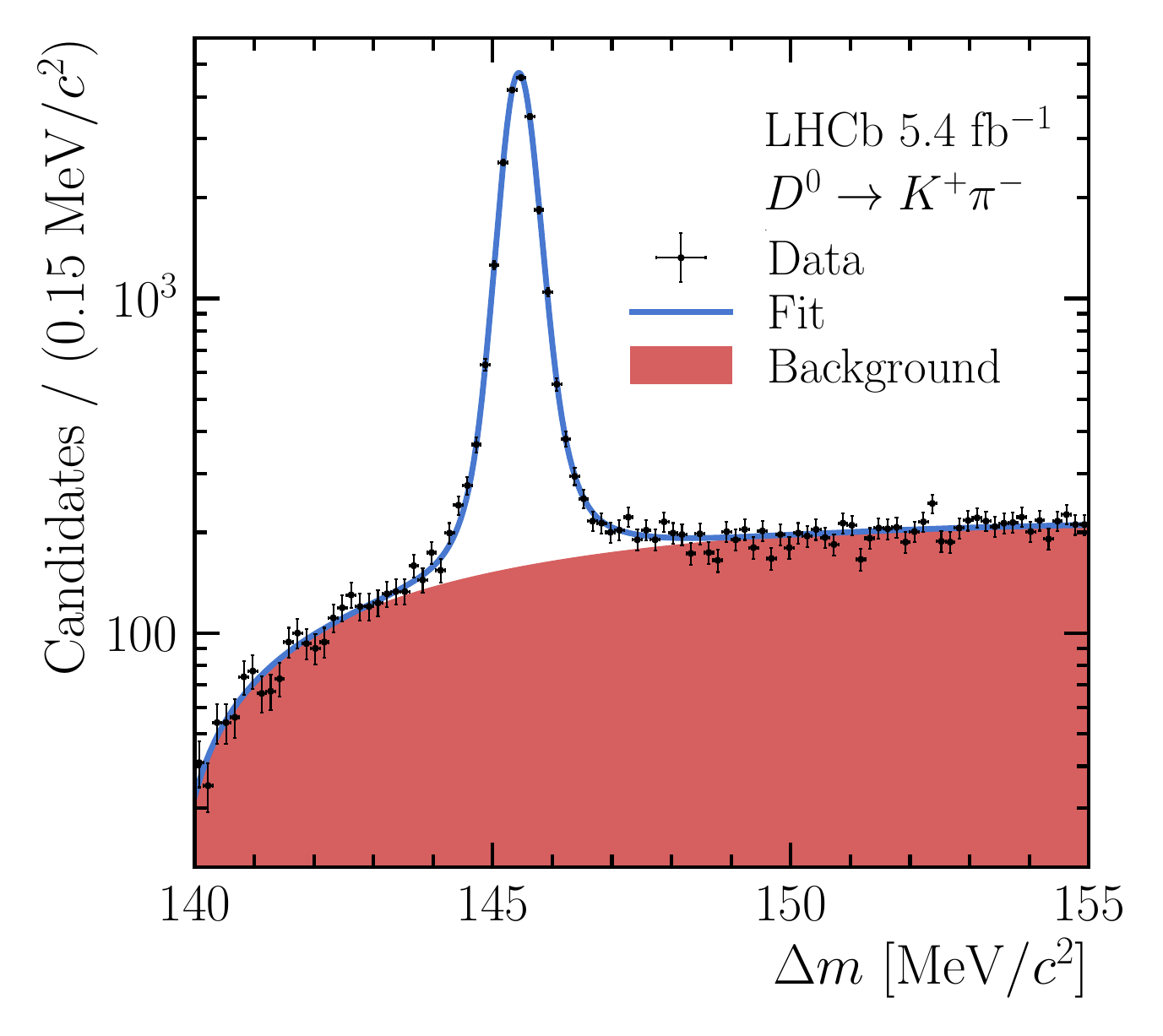}
     \end{subfigure}
    \caption{Distribution of the mass difference, $\Delta m$, between the $\Dstarp$ and $\Dz$ mesons in (left) RS and (right) WS samples, with the result of the extended maximum-likelihood fit performed simultaneously on both samples also shown.}
    \label{fig:selection/dM}
\end{figure}

Following candidate selection the RS and WS signal yields are approximately $5.2\times10^{6}$ and $2.0\times10^{4}$, respectively.
These are obtained from an extended unbinned maximum-likelihood fit to the \dm distribution.
The distribution, shown in Fig.~\ref{fig:selection/dM}, has an eight-fold improved resolution with respect to the masses of the \Dstarp or \Dz individually.
It is described by a combination of probability density functions (PDFs) to model the signal and an ARGUS~\cite{Argus} PDF to model the combinatorial background.
The signal model consists of a Johnson $S_U$ PDF~\cite{Johnson1,Johnson2} and three Gaussian PDFs, one of which has asymmetric widths.
The signal model is used to describe both RS and WS data simultaneously while independent ARGUS PDFs model the combinatorial background in each.

\section{Measurement strategy}
\label{sec:ratio}

The dataset is divided according to the measured \Dz decay time into eight subsamples.
Division boundaries are chosen ensuring each subsample contains an approximately equal RS yield while the minimum bin width is constrained to be larger than the decay-time resolution of the \lhcb detector.
This resolution, determined by fitting the RS decay-time distribution with the convolution of an exponential PDF with three Gaussian functions, is measured to be $78\fs$.
The eight subsamples are then further divided according to the flavour, \Dz or \Dzb, using the charges of the muon and soft pion tagging particles.
The RS and WS yields in each subsample are extracted from \dm extended maximum-likelihood fits using the same model as that described previously.
Initial values of the model parameters in the extended likelihood maximisation are set to those obtained from the fit to the entire dataset.
All are allowed to vary except the tail parameters of the Johnson $S_U$ PDF which are fixed globally to ensure convergence.

The quadratic mixing approximation defined by Eq.~\ref{eqn: quadratic mixing approximation} and its relation to the decay-rate ratios in Eqs.~\ref{eqn: R+}~and~\ref{eqn: R-} are discretised to accommodate the sixteen subsamples  corresponding to decay-time bin $i$ and to \Dz (\Dzb) flavour denoted by $+$ ($-$):
\begin{equation}
    R_i^\pm  = \RDpm + \sqrt{\RDpm}\ypm\frac{\langle{t}\rangle_i^\pm}{\tau_{\Dz}} + \frac{\xsqpm+(\ypm)^{2}}{4}\frac{\langle{t^2}\rangle_i^\pm}{\tau_\Dz^2}.
    \label{eqn: uncorrected ratio function}
\end{equation}
The continuous lifetime $t$ and its square are replaced with the mean time $\langle t\rangle$ and mean-square time $\langle t^2\rangle$ per subsample.
Consequently, the decay-rate ratios are substituted by simple ratios, $r_i^\pm$, of the number of decays that have been tagged as WS, $N_{\text{WS},i}^\pm$, and RS, $N_{\text{RS},i}^\pm$, in each subsample.
Values of $\langle{t^{(2)}}\rangle_i^\pm$ are calculated from the RS subsamples with the background subtracted using the \sPlot technique~\cite{Pivk:2004ty}.

While the periodic reversal of the magnet polarity minimises detection efficiency asymmetries, there are residual effects originating from the differing interactions of certain particles and their antiparticle counterparts with the detector.
Similarly, there is an asymmetry in the production rates of \B and \Bb mesons at the \lhc.
Taking the WS-to-RS ratio exactly cancels the \Bb meson production asymmetry and the detection asymmetries of the tagging particles, since these particles are the same in both WS and RS samples.
Only the asymmetry in the detection efficiencies of the \Dz decay products, $A_{\kaon\pion}$, remains.
This is defined by the detection efficiencies of each final state as
\begin{equation}
    A_{\kaon\pion} = \frac{\varepsilon_{\Km\pip} - \varepsilon_{\Kp\pim}}{\varepsilon_{\Km\pip} + \varepsilon_{\Kp\pim}}.
\end{equation}
The detection asymmetry is used to correct $R_i^\pm$ along with an additional correction for misidentified candidates migrating from the RS to the WS sample which results in shifts, $b_i^\pm$, in the measured ratios.
The ratio corrected for these biasing nuisance parameters measured from data is then
\begin{equation}
    \tilde{R}_i^\pm = \frac{1\mp A_{\kaon\pion}}{1\pm A_{\kaon\pion}} R_i^\pm + b_i^\pm.
    \label{eqn:corrected ratio}
\end{equation}

\section{Determination of nuisance parameters}

\subsection{Particle detection asymmetry}
\label{sec: asymmetry}

\begin{figure}
    \centering
     \includegraphics[width=0.6\linewidth]{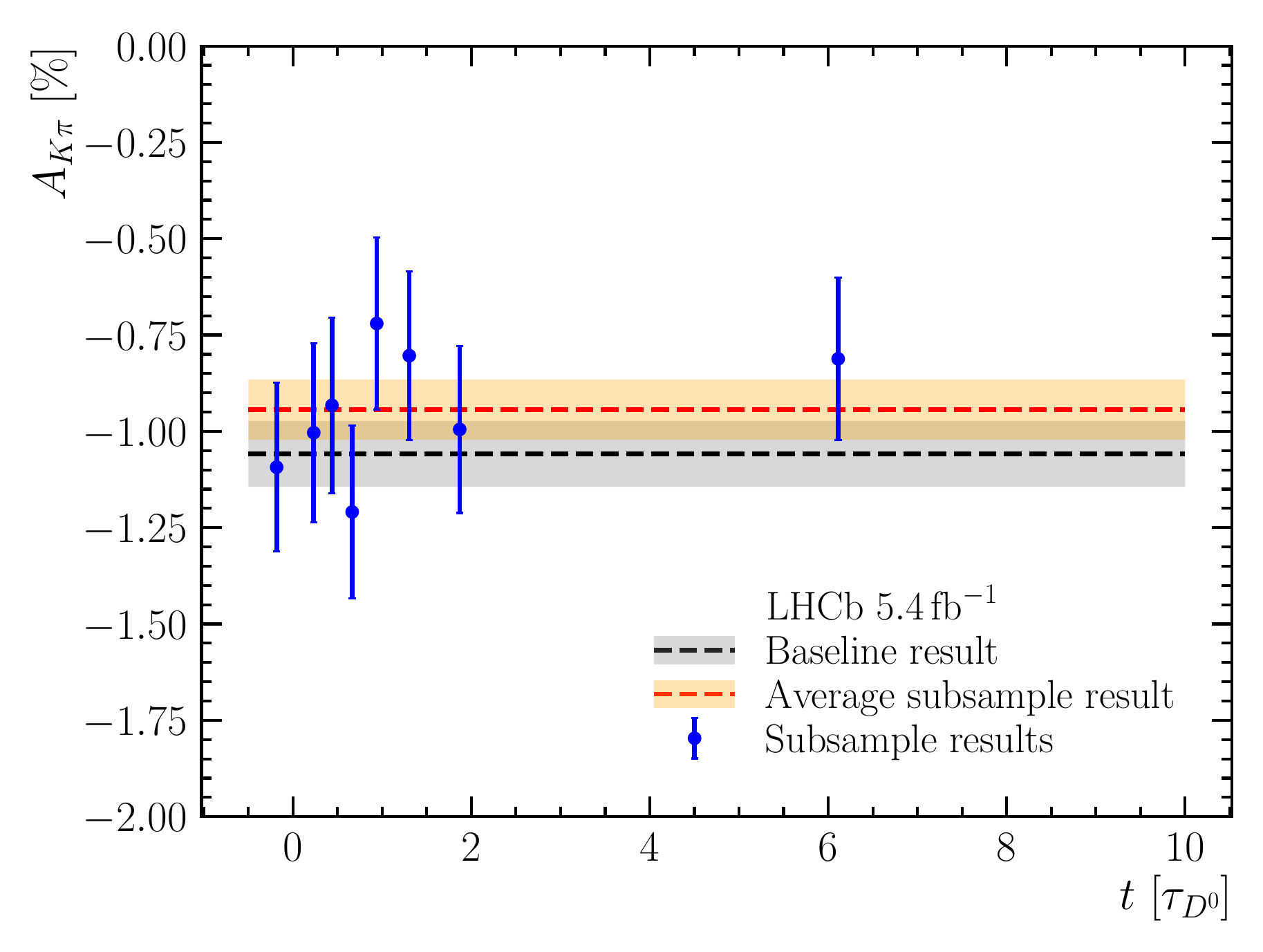}
    \caption{Measurements of the instrumental asymmetry, $A_{K\pi}$, in each of the decay-time subsamples together with their average and the baseline result across all data.}
    \label{fig:asymmetry}
\end{figure}

The detection asymmetry $A_{\kaon\pion}$ is measured from prompt ${\Dpm\to\Kmp\pipm\pipm}$ and ${\Dpm\to\KS\pipm}$ calibration data and uses the \KS detection asymmetry, $A_\KS^d$, as an external input from Ref.~\cite{LHCb-PAPER-2013-003}.
It is given by
\begin{equation}
    A_{\kaon\pion} = A_{\kaon\pion\pion}^{\text{raw}} - A_{\KS\pion}^{\text{raw}} + A_\KS^d,
\end{equation}
where the raw asymmetries,
\begin{equation}
    A^{\text{raw}} = \frac{N_\Dp-N_\Dm}{N_\Dp+N_\Dm},
\end{equation}
are calculated from the numbers of \Dpm decays, $N_\Dpm$.
In this case the raw asymmetries are directly obtained from extended \chisq fits to the invariant-mass distributions of the \Dpm candidates.
Prior to fitting, a weighting~\cite{Rogozhnikov:2016bdp} of the calibration data is performed such that the asymmetry, which is measured indirectly from calibration data, is representative of the signal sample.
The ${\Dpm\to\Kmp\pipm\pipm}$ dataset is weighted first, using the kinematics of the kaons and the overlapping pions, to the signal data.
The overlapping pions are the pion of the \Dz decay in signal data and the pion with the lower transverse momentum in the ${\Dpm\to\Kmp\pipm\pipm}$ data, whose kinematical distributions overlap most closely.
The resulting weighted kinematics of the \Dpm meson and the leftover pion are in turn used to weight the ${\Dpm\to\KS\pipm}$ dataset.
The external measurement of $A_\KS^d$ is scaled according to the changes in the \KS kinematic distributions that are induced by applying the generated weights.
Since the detection of charged kaons and pions is independent of their decay ancestry, $A_{\kaon\pion}$ is not expected to differ for \Dz and \Dzb decays.
Dependence on the decay lifetime is also not expected.
This is verified by repeating the measurement of $A_{\kaon\pion}$ in the eight decay-time subsamples.
Each is plotted in Fig.~\ref{fig:asymmetry} along with their average and the baseline decay-time-integrated result, the difference between which is added to the statistical uncertainty in quadrature.
The resulting instrumental asymmetry is ${A_{\kaon\pion} = (-1.06 \pm 0.14)\,\%}$, where the uncertainty contains both the statistical uncertainty and the uncertainties derived from the described systematic effects.

\subsection{Misidentified backgrounds}
\label{sec: peaking}

Selected candidates obtained by combining misidentified final-state particles can also bias the measurements.
The restrictions on $m(\kaon\pion)$ completely suppress contamination from ${\Dz\to\Kp\Km}$ and ${\Dz\to\pip\pim}$ decays.
Double-misidentified ${\Dz\to\Kpm\pimp}$ decays, where the pion is identified as a kaon and vice-versa, however, remain.
Such misidentifications contaminate the WS sample with RS decays and therefore bias the subsample ratios; the reverse is also possible but has negligible impact due to the relative sizes of the WS and RS samples.

\begin{figure}
    \centering
     \includegraphics[width=\textwidth]{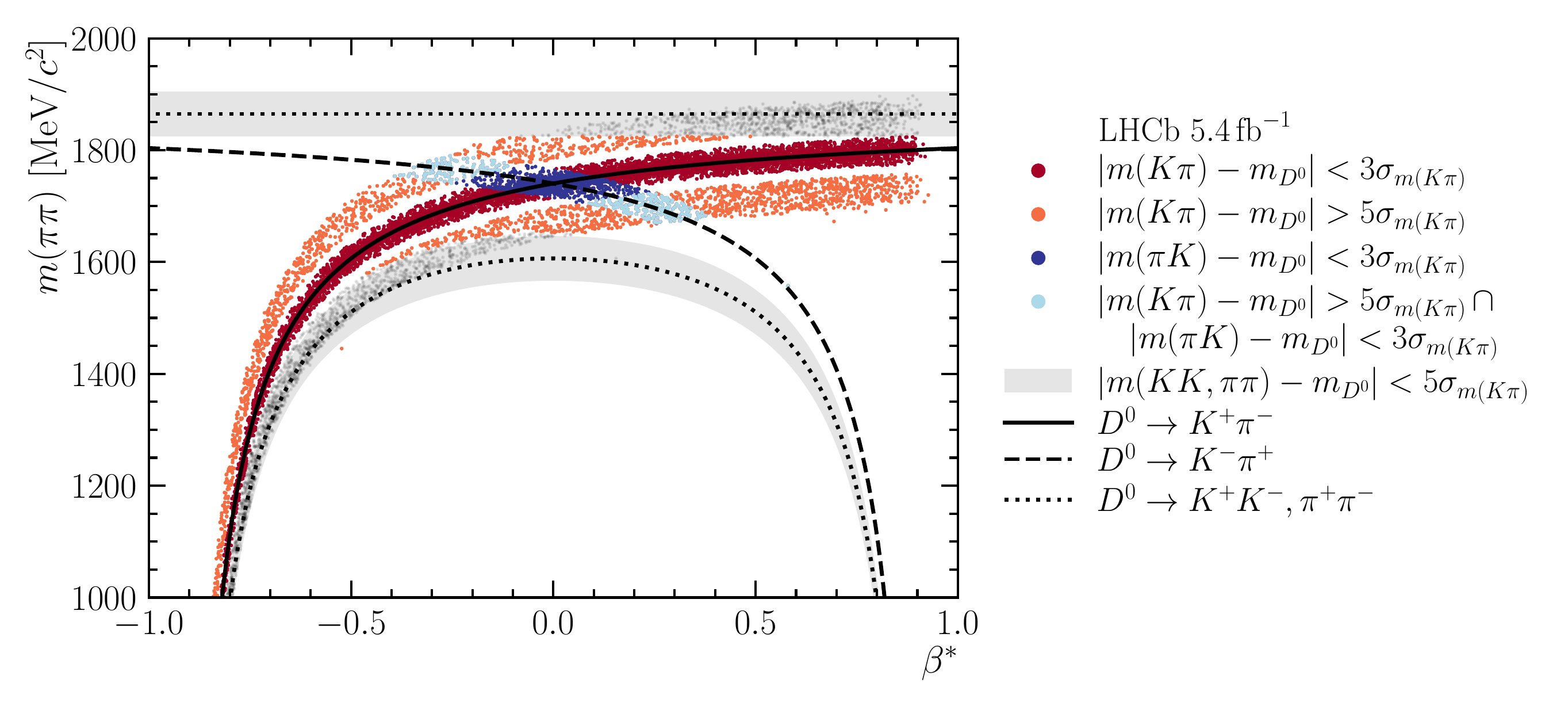}
    \caption{Distribution of WS candidates in a two-dimensional phase space, where ${m(KK,\pi\pi)}$ is the $\Dz$ invariant mass under the ${\Dz\to\Kp\Km,\pi^+\pi^-}$ hypothesis and $\beta^{*}$ is the momentum imbalance of the $\Dz$ decay particles. The $\Dz$ signal and mass sideband subsets are illustrated by the red and orange regions, respectively. Analytical solutions are also shown as black lines along with the regions defined by the veto on ${m(KK,\pi\pi)}$ shaded in grey. Contamination from RS decays in the signal and mass sideband subsets is illustrated by the dark and light blue regions, respectively.}
    \label{fig:beta}
\end{figure}

\begin{figure}
    \centering
    \includegraphics[width=0.45\textwidth]{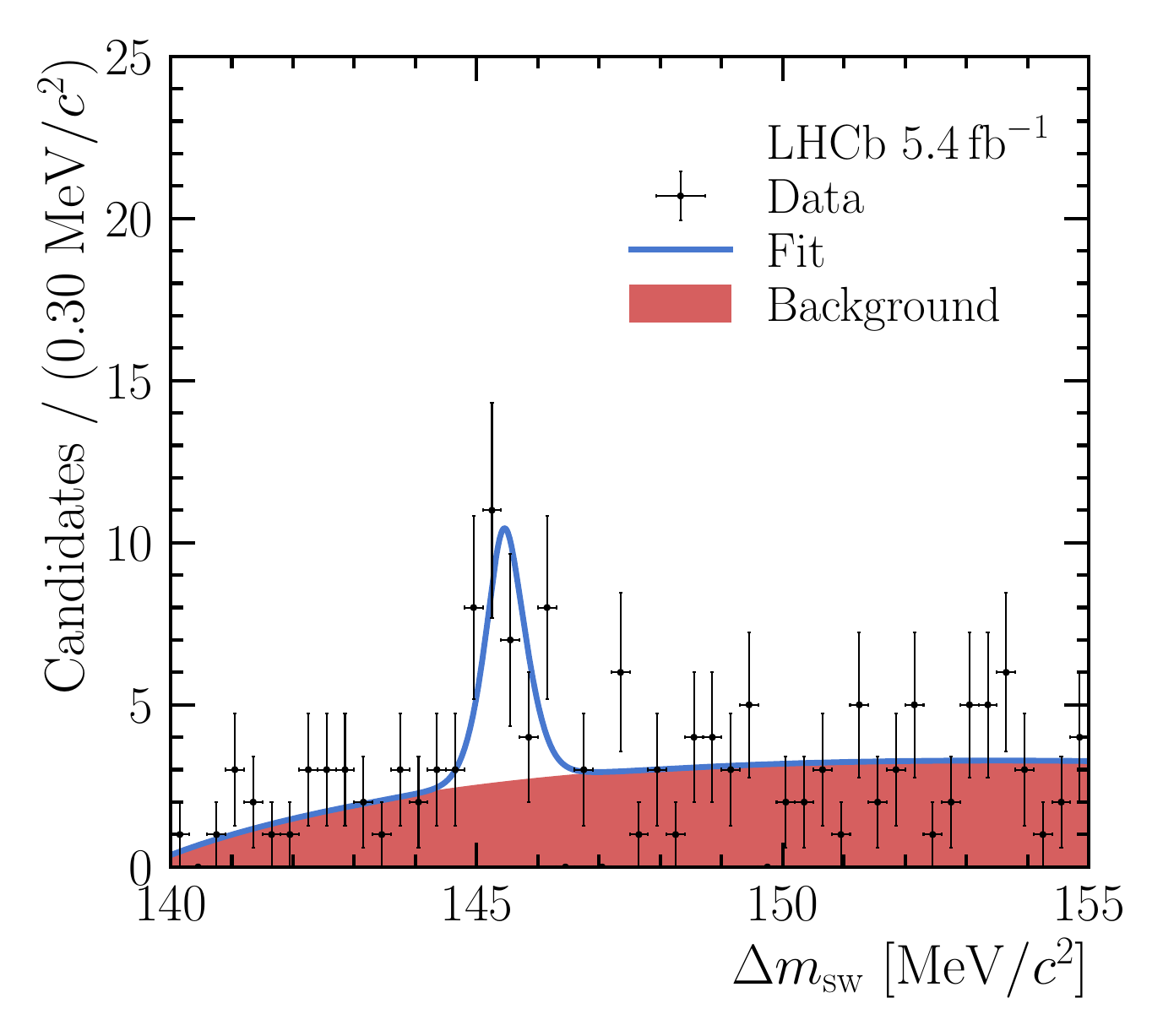}
    \includegraphics[width=0.45\textwidth]{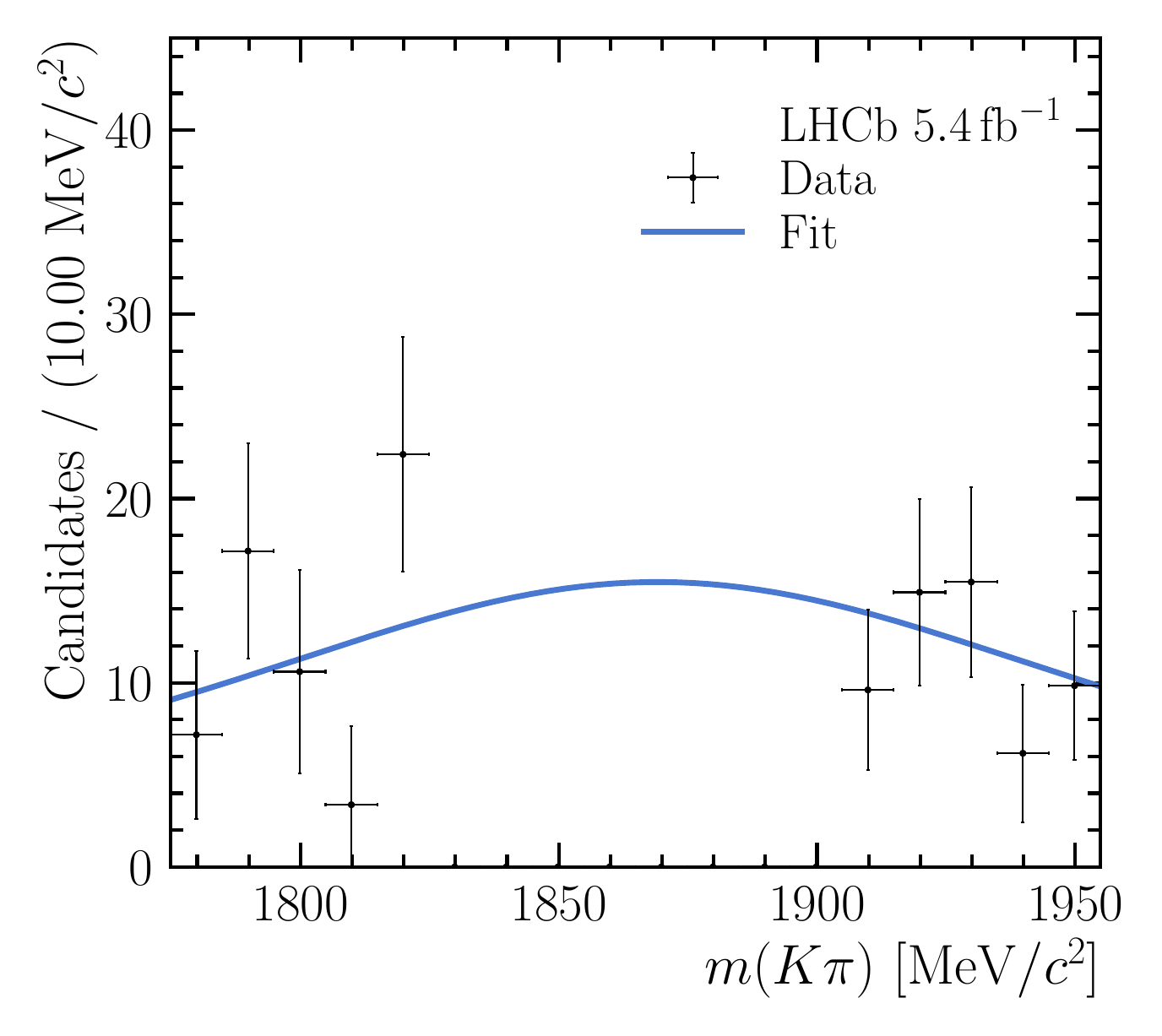}
    \caption{(Left) An example of the $\Delta m_{\text{sw}}$ distribution and the corresponding fit result in one sideband interval, ${[m_\Dz-50 \, \text{Me}\kern -0.1em\text{V}\!/c^2, m_\Dz-40 \, \text{Me}\kern -0.1em\text{V}\!/c^2]}$. (Right) Measured yield of double-misidentified decay candidates in each interval with the fit result also shown.}
    \label{fig:peaking}
\end{figure}

The number of double-misidentified decays which peak in the signal region ${|m(\kaon\pion)-m_\Dz|<3\sigma_{m(\kaon\pion)}}$ in the WS sample, $N_{\pion\kaon}$, is measured by recalculating \dm after swapping the \kaon and \pion mass hypotheses, denoted by $\dm_{\text{sw}}$.
Since it is RS decays that form this background in WS data, the selection requirement that vetoes data populating the region defined by ${|m(\kaon\pion)-m_\Dz|>3\sigma_{m(\kaon\pion)}}$ is replaced by an equivalent requirement on the swapped mass hypothesis invariant mass, $m(\pion\kaon)$.
Directly measuring the number of double-misidentified decays in this region would lead to an overestimation, since the ${|m(\kaon\pion,\pion\kaon)-m_\Dz|<3\sigma_{m(\kaon\pion)}}$ regions overlap.
This is illustrated by the crossing of the signal (red) and swapped mass (dark and light blue) branches in WS data in Fig.~\ref{fig:beta}.
The figure plots $\beta^{*}$, a metric of the momentum imbalance between the $\Dz$ decay particles, against the $\Dz$ invariant mass calculated under the ${\Dz\to\pip\pim}$ decay hypothesis, $m(\pion\pion)$. The momentum imbalance is defined as
\begin{equation}
    \beta^{*} = q(\spion)\frac{|\vec{p}_{+}|-|\vec{p}_{-}|}{|\vec{p}_{+}|+|\vec{p}_{-}|},
    \label{eqn: beta *}
\end{equation}
where $\vec{p}_{+}$ and $\vec{p}_{-}$ are the momenta of the positively and negatively charged decay particles, respectively, and $q(\spion)$ is the charge on the soft pion.
Analytical solutions for each considered decay of the \Dz are also drawn in the figure as black lines.
The number of double-misidentified decays in the WS sample is therefore measured using candidates in the light blue regions either side of the central signal strip.
This is defined simultaneously by ${|m(\kaon\pion)-m_\Dz|>5\sigma_{m(\kaon\pion)}}$, the so-called sidebands, and ${|m(\pion\kaon)-m_\Dz|<3\sigma_{m(\kaon\pion)}}$.
The $\dm_{\text{sw}}$ distribution is fitted, similarly to the fits to $\dm$ for signal yield extraction, to extract the number of double-misidentified decays.
The signal is modelled by the sum of a Johnson $S_U$ PDF and a Gaussian PDF, with an example shown in Fig.~\ref{fig:peaking}~(left).
The measured yields in the sidebands are subsequently interpolated to measure the yield in the dark blue signal region of Fig.~\ref{fig:beta}.

Figure~\ref{fig:peaking}~(right) shows the determined yield of double-misidentified decays in ten intervals divided evenly across both $m(\kaon\pion)$ sidebands.
The yield across the entire $m(\kaon\pion)$ range is scaled to the signal region within three standard deviations of the $\Dz$ invariant mass. 
This scaled yield is measured as ${N_{\pion\kaon} = 72 \pm 12}$.
The swapped mass sample size is not large enough to measure the number $N_{\pion\kaon,i}^{\pm}$ in each subsample.
This is instead estimated with RS simulation where the fraction of events in the $\dm_{\text{sw}}$ distribution of each simulated subsample is used to scale the total yield $N_{\pion\kaon}$ to the corresponding subsample of data.
The shifts in Eq.~\ref{eqn:corrected ratio} are then ${b_{i}^{\pm} = N_{\pion\kaon,i}^{\pm} / N_{\text{RS},i}^{\pm}}$ with values of the order of ${10^{-5}}$.

\section{Measurement of mixing and \CP violation}

The flavour oscillation variables $\RDpm$, $\ypm$ and $\xsqpm$ are determined by minimising the \chisq function
\begin{equation}
    \chisq = \sum_{i,q} \left[ \left( \frac{r_i^q - \tilde{R}_i^q}{\sigma_{r_i^q}} \right)^2 + \chisq_{\text{nuis}}(b_i^q)\right] + \chisq_{\text{nuis}}(A_{\kaon\pion}),
    \label{eqn:chi2function}
\end{equation}
where the indices $i$ and $q$ iterate over the eight decay-time subsamples and two meson flavours, respectively.
The uncertainty on the ratios, $\sigma_{r_i^q}$, is propagated from the simultaneous measurement of RS and WS yields,
\begin{equation}
    \left(\sigma_{r_i^q}\right)^2 = \left(\frac{\sigma_{N_{\text{RS},i}^q}}{N_{\text{RS},i}^q}\right)^2 + \left(\frac{\sigma_{N_{\text{WS},i}^q}}{N_{\text{WS},i}^q}\right)^2 - \frac{2\sigma_{N_{\text{RS},i}^qN_{\text{WS},i}^q}}{N_{\text{RS},i}^qN_{\text{WS},i}^q},
\end{equation}
where $\sigma_{N_{\text{RS},i}^qN_{\text{WS},i}^q}$ is the covariance between the RS and WS yields, while $\sigma_{N_{\text{RS},i}^q}$ and $\sigma_{N_{\text{WS},i}^q}$ are their respective uncertainties.
The $\chisq_{\text{nuis}}(\Xi)$ terms constrain the corresponding nuisance parameter $\Xi\in[A_{\kaon\pion},b_{i}^{\pm}]$, that is defined in $\tilde{R}_i^q$ along with the variables of interest, to the corresponding measured value ${\xi\pm\sigma_\xi}$,
\begin{equation}
    \chisq_{\text{nuis}} = \left(\frac{\xi-\Xi}{\sigma_\xi}\right)^2.
\end{equation}
The measured ratios $r_i^q$ are determined from the Run~2 dataset for the baseline result.

\subsection{Consistency checks}
Several consistency checks are performed on disjoint subsets of selected candidates chosen to expose potentially neglected sources of systematic uncertainty.
Subsets divide the data according to data-taking year, orientation of the magnetic field, transverse momenta of both tagging particles and momentum of the kaon.
The \chisq difference of each parameter ${\zeta=[\RDpm,\ypm,\xsqpm]}$ with respect to the baseline fitted value $\zeta_{\text{base}}$,
\begin{equation}
    \chisq_\zeta = \sum_{j}\left(\frac{\zeta_j-\zeta_{\text{base}}}{\sigma_{\zeta_j}}\right)^2,
\end{equation}
is calculated for each of the consistency checks, where $j$ iterates over the subsets defined specifically for each check.
The corresponding $p$-value, defined by ${p_\zeta=1-F_{\chisq}(\chisq_\zeta, n_j-1)}$, where $F_{\chisq}(z,k)$ is the cumulative distribution function of the \chisq distribution with $k$ degrees of freedom evaluated at $z$, is determined for each parameter and for each check.
If no biases are present, the distribution of many such $p$-values is uniform.
The nonuniformity of the 48
(6 parameters in each of 8 consistency checks) values of $p_\zeta$ is quantified by a Kolmogorov--Smirnov test~\cite{Kolmogorov,Smirnov}.
The resulting $p$-value of $60.7\,\%$ indicates agreement with a uniform distribution.

\subsection{Systematic uncertainties}
In addition to the statistical uncertainties obtained from the decay-time fit, several sources of systematic uncertainty are considered.
These are summarised in Table~\ref{tab:systematics} for the case of the decay-time fit allowing \CP violation both in mixing and decay.
The totals are given by the corresponding quadrature sum for each of the fit parameters.
Analogous systematic uncertainties are determined for the parameters measured under the remaining two hypotheses.

\begin{table}
    \centering
    \caption{Identified sources of systematic uncertainty associated with each parameter of the decay-time fit allowing \CP violation both in mixing and decay.}
    \label{tab:systematics}
    
    \begin{tabular}{ c  c c c c c c }
        \hline
         Source & \makecell{$\RDp$ \\ $[10^{-5}]$} & \makecell{$\yp$ \\ $[10^{-3}]$} & \makecell{$\xsqp$ \\ $[10^{-4}]$} & \makecell{$\RDm$ \\ $[10^{-5}]$} & \makecell{$\ym$ \\ $[10^{-3}]$} & \makecell{$\xsqm$ \\ $[10^{-4}]$} \\
        \hline
         \dm                       & $0.75$ & $0.21$ & $0.11$ & $1.08$ & $0.25$ & $0.11$ \\
         $\langle{t^{(2)}}\rangle$ & $0.02$ & $0.02$ & $0.01$ & $0.02$ & $0.01$ & $0.04$ \\
         $A_{\kaon\pion}$          & $1.75$ & $0.00$ & $0.01$ & $1.63$ & $0.03$ & $0.01$ \\
         $N_{\pion\kaon}$          & $0.22$ & $0.00$ & $0.01$ & $0.22$ & $0.00$ & $0.00$ \\
         $f_{\text{prompt}}$       & $1.17$ & $0.10$ & $0.03$ & $1.12$ & $0.10$ & $0.04$ \\
         $\Delta\mu_{\text{fit}}$  & $0.37$ & $0.09$ & $0.02$ & $0.39$ & $0.11$ & $0.05$ \\
        \hline
        Total & $2.28$ & $0.25$ & $0.12$ & $2.30$ & $0.29$ & $0.13$ \\
        \hline
    \end{tabular}
\end{table}

The signal yields, determined from extended maximum-likelihood fits to \dm distributions, ultimately depend on the PDFs empirically chosen to model the \dm distributions.
Different yields are obtained when fitted with an alternative model, in turn shifting the ratios and the fitted parameters of interest.
Many possible PDF combinations were considered but most do not describe the \dm distribution well, particularly in the larger RS dataset.
The next best model consists of an ARGUS PDF, a Johnson $S_U$ PDF and three Gaussian PDFs which share a mean parameter.
It differs from the baseline model by replacing the Gaussian PDF with asymmetric widths by a Gaussian PDF which shares a mean with the other two.
The extent to which using the alternative model changes the measured value of each parameter is evaluated by randomly generating many pseudoexperiments according to the fit results of the alternative model to the \dm distribution of each subsample.
The distributions of the generated pseudoexperiments are fitted with both models and the resulting yields propagated to the decay-time fits that are performed for each model.
The systematic uncertainties are calculated as the quadrature sum of the mean and standard deviation of the distribution of differences between the fitted results from each model, for each parameter.

Values of the average time and time-squared, $\langle{t^{(2)}}\rangle$, in each subsample are sensitive to the migration of candidates between the subsamples.
A conservative estimate of this effect is evaluated with the help of pseudoexperiments, in which the smearing of the decay time of the simulated events is applied or not.
The decay-time fits are performed on pseudoexperiments and the systematic uncertainty is determined by looking at the distributions of results with smearing applied and not, taking the quadrature sum of the differences in the mean and its uncertainty.

The results are tested for sensitivity to variations in the fit nuisance parameters that are associated with the instrumental asymmetry $A_{\kaon\pion}$ and the number of double-misidentified decays in the WS sample $N_{\pion\kaon}$.
For both, the tests are three-fold: the measured values of the nuisance parameters are increased by one standard deviation, decreased by one standard deviation, and fixed to their central value.
The maximum deviation from the baseline value across all three tests is assigned as the systematic uncertainty for each fit parameter.

The yield of prompt \Dz candidates is expected to be negligible in the signal dataset as a result of triggering only on events containing muons.
The residual fraction $f_{\text{prompt}}$ in each decay-time subsample is estimated by fitting the $\ln(\chisqip)$ distribution of the \Dz meson, where contributions from prompt and secondary decays are well separated.
The fraction is very small, $<0.5\aunit{\%}$, and therefore difficult to extract in RS data and not possible in the smaller WS dataset.
A systematic uncertainty is assigned for the WS subsamples by taking two extreme cases: modifying the measured fraction in the corresponding RS subsample by taking either half of it, or doubling it.
The ratios are then scaled to either extreme case before reprocessing the decay-time fits and assigning the maximum deviation of each fit parameter from the baseline values as a systematic uncertainty.

Finally, differences are observed between the input oscillation parameters used for the generation of pseudoexperiments and their output values from the subsequent decay-time fits.
A systematic uncertainty $\Delta\mu_{\text{fit}}$ is found from the pull distribution of many pseudoexperiments, where the pull is defined as the difference divided by the statistical uncertainty of the fitted pseudoresult.
As before, the quadrature sum of the observed bias and its uncertainty is used.

\section{Results}
\label{sec: Results}

\begin{figure}
    \centering
     \includegraphics[width=0.7\textwidth]{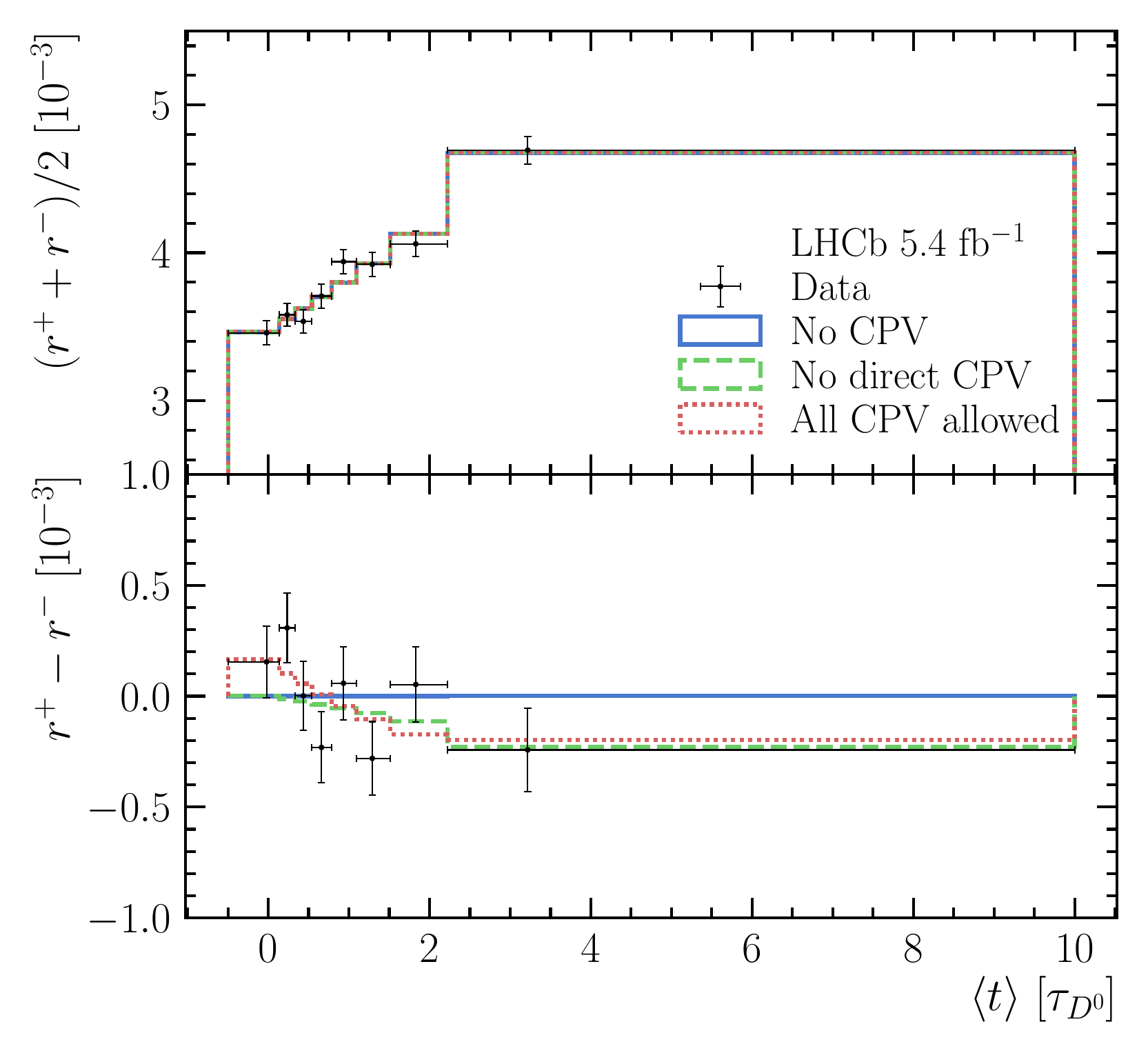}
    \caption{(Top) Average of the measured WS-to-RS decay ratios of $\Dz$ and $\ensuremath{{\ensuremath{\ensuremath{\kern 0.18em \overline{\kern -0.18em \PD}}}}\xspace{}^0}\xspace$ mesons and (bottom) their difference, as a function of the average decay time $\langle{t}\rangle$.
    Vertical and horizontal error bars correspond to the statistical uncertainty $\sigma_{r^{\pm}}$ and decay-time bin boundaries, respectively. Projections of the fit hypotheses are also shown.}
    \label{fig:result}
\end{figure}

For the final result three fit hypotheses are considered.
The primary hypothesis is tested via the decay-time fit allowing \CP violation both in mixing and decay and is described by Eq.~\ref{eqn: uncorrected ratio function}.
Next, there is no direct \CP violation if, like RS decays, the WS decays of \Dz and \Dzb mesons are \CP symmetric, ${|\mathcal{A}_{\Kp\pim}|=|\overline{\mathcal{A}}_{\Km\pip}|}$.
This is tested by fixing ${\RDp=\RDm\equiv\RD}$, subsequently removing one degree of freedom.
Lastly, we test the case of no \CP violation by additionally fixing ${\yp=\ym\equiv\y}$ and ${\xsqp=\xsqm\equiv\xsq}$.
This corresponds to a measurement of the $\RD$, $\y$ and $\xsq$ parameters only.
Fit projections under all hypotheses are shown in Fig.~\ref{fig:result}.

Table~\ref{tab:result_ACPVA} presents the result of the decay-time fit allowing \CP violation both in mixing and decay, and the correlations between the parameters. 
The quoted uncertainties are the quadrature sums of the statistical and systematic components and are statistically dominated.
The correlations incorporate the statistical correlations obtained from the decay-time fit and the systematic contributions.
The \CP asymmetry in the decay, defined by
\begin{equation}
    \AD \equiv \frac{\RDp-\RDm}{\RDp+\RDm},
\end{equation}
is measured to be ${\AD=(2.3\pm1.7)\,\%}$.
Table~\ref{tab:result_NDCPV} shows the results for the fit where $\RDp$ and $\RDm$ are constrained to be identical.
Table~\ref{tab:result_MO} shows the result of the mixing-only fit.
All results are consistent with \CP symmetry.
The mixing parameters defined in Eq.~\ref{eqn:x'y'+-} are determined simultaneously on both the $\Dz$ and $\Dzb$ subsamples from the fit performed under the no \CP violation hypothesis to be ${(x^\prime)^2=(0.0\pm1.2)\times10^{-4}}$ and ${y^\prime=(5.8\pm1.6)\times10^{-3}}$.

\begin{table}[tb]
    \centering
    \caption{Result of the decay-time fit allowing \CP violation both in mixing and decay. The first uncertainty is statistical and the second systematic. Correlations include both statistical and systematic contributions.}
    \begin{tabular}{ c R | S[table-format=3.1] S[table-format=3.1] S[table-format=3.1] S[table-format=3.1] S[table-format=3.1] S[table-format=3.1] }
        \hline
        \multirow{2}{*}{Parameter} & \multicolumn{1}{c}{\multirow{2}{*}{Value}} & \multicolumn{6}{c}{Correlations [\%]} \\
        \multicolumn{2}{c}{} & $\RDp$ & $\yp$ & $\xsqp$ & $\RDm$ & $\ym$ & ${(x^{\prime-})}^{2}$ \\
        \hline
        $\RDp$  & $( 355.2 \pm 7.9 \pm 2.3) \times 10^{-5}$ &  100.0 &  -76.2 &   63.4 &   -1.5 &   -0.1 &    0.0 \\
        $\yp$   & $(   3.6 \pm 2.2 \pm 0.3) \times 10^{-3}$ &        &  100.0 &  -94.2 &   -0.0 &   -0.0 &    0.0 \\
        $\xsqp$ & $(   1.1 \pm 1.6 \pm 0.1) \times 10^{-4}$ &        &        &  100.0 &   -0.0 &   -0.0 &    0.0 \\
        $\RDm$  & $( 339.1 \pm 7.9 \pm 2.3) \times 10^{-5}$ &        &        &        &  100.0 &  -76.2 &   64.6 \\
        $\ym$   & $(   8.1 \pm 2.3 \pm 0.3) \times 10^{-3}$ &        &        &        &        &  100.0 &  -94.9 \\
        $\xsqm$ & $(  -1.1 \pm 1.9 \pm 0.1) \times 10^{-4}$ &        &        &        &        &        &  100.0 \\
        \hline
    \end{tabular}
    \label{tab:result_ACPVA}
\end{table}

\begin{table}[t]
    \centering
    \caption{Result of the no direct \CP violation fit. The first uncertainty is statistical and the second systematic. Correlations include both statistical and systematic contributions.}
    \begin{tabular}{ c R | S[table-format=3.1] S[table-format=3.1] S[table-format=3.1] S[table-format=3.1] S[table-format=3.1] }
        \hline
        \multirow{2}{*}{Parameter} & \multicolumn{1}{c}{\multirow{2}{*}{Value}} & \multicolumn{5}{c}{Correlations [\%]} \\
        \multicolumn{2}{c}{} & $\RD$ & $\yp$ & $\xsqp$ & $\ym$ & $\xsqm$ \\
        \hline
        $\RD$   & $( 347.1 \pm 5.5 \pm 1.3) \times 10^{-5}$ &  100.0 &  -65.0 &   51.6 &  -65.0 &   51.5 \\
        $\yp$   & $(   5.4 \pm 1.9 \pm 0.5) \times 10^{-3}$ &        &  100.0 &  -90.1 &   43.2 &  -34.3 \\
        $\xsqp$ & $(   0.0 \pm 1.5 \pm 0.3) \times 10^{-4}$ &        &        &  100.0 &  -34.3 &   27.3 \\
        $\ym$   & $(   6.3 \pm 1.9 \pm 0.4) \times 10^{-3}$ &        &        &        &  100.0 &  -90.9 \\
        $\xsqm$ & $(   0.1 \pm 1.6 \pm 0.3) \times 10^{-4}$ &        &        &        &        &  100.0 \\
        \hline
    \end{tabular}
    \label{tab:result_NDCPV}
\end{table}

\begin{table}[t]
    \centering
    \caption{Result of the no \CP violation fit. The first uncertainty is statistical and the second systematic. Correlations include both statistical and systematic contributions.}
    \begin{tabular}{ c R | S[table-format=3.1] S[table-format=3.1] S[table-format=3.1] }
        \hline
        \multirow{2}{*}{Parameter} & \multicolumn{1}{c}{\multirow{2}{*}{Value}} & \multicolumn{3}{c}{Correlations [\%]} \\
        \multicolumn{2}{c}{} & $\RD$ & $y^{\prime}$ & ${(x^{\prime})}^{2}$ \\
        \hline
        $\RD$              & $( 347.0 \pm 5.5 \pm 1.4) \times 10^{-5}$ &  100.0 &  -77.9 &   65.2 \\
        $y^{\prime}$       & $(   5.8 \pm 1.6 \pm 0.2) \times 10^{-3}$ &        &  100.0 &  -94.7 \\
        $(x^{\prime})^{2}$ & $(   0.0 \pm 1.2 \pm 0.1) \times 10^{-4}$ &        &        &  100.0 \\
        \hline
    \end{tabular}
    \label{tab:result_MO}
\end{table}

\begin{figure}
    \centering
    \includegraphics[width=0.45\textwidth]{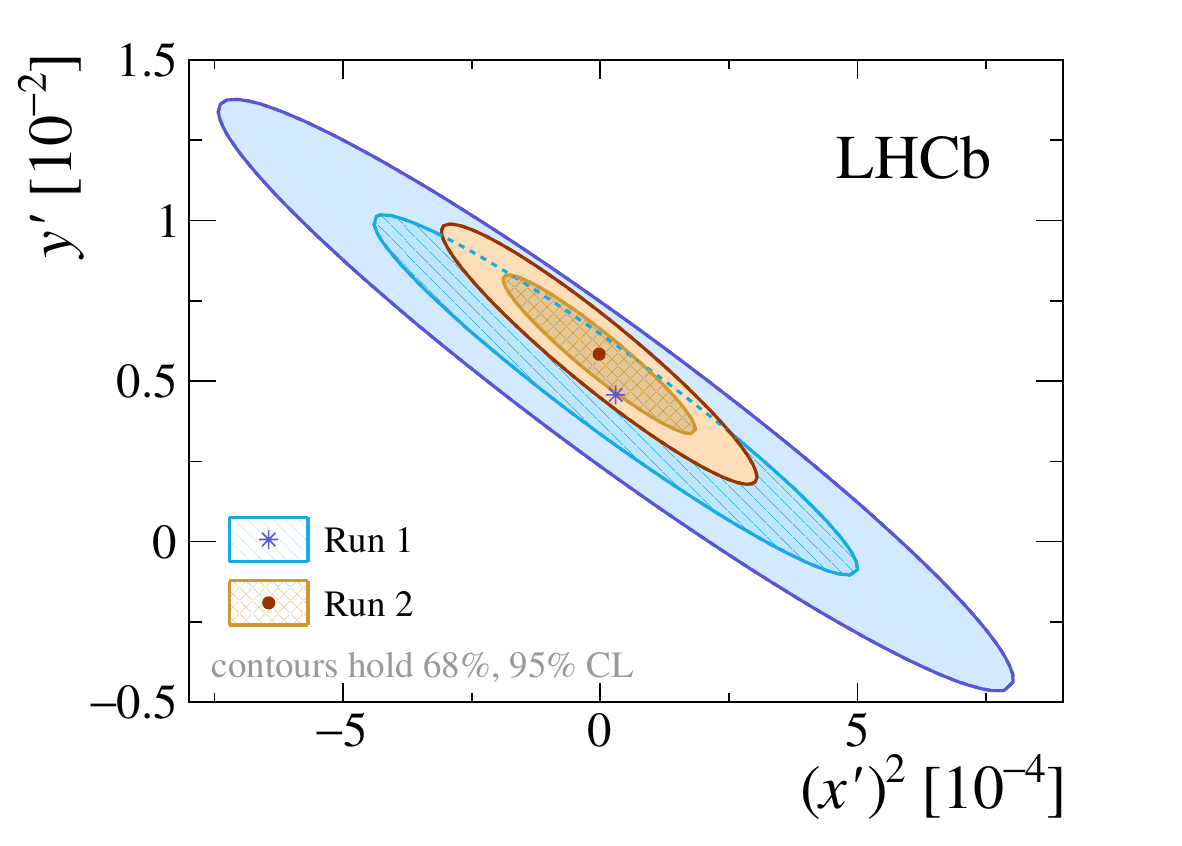}
    \includegraphics[width=0.45\textwidth]{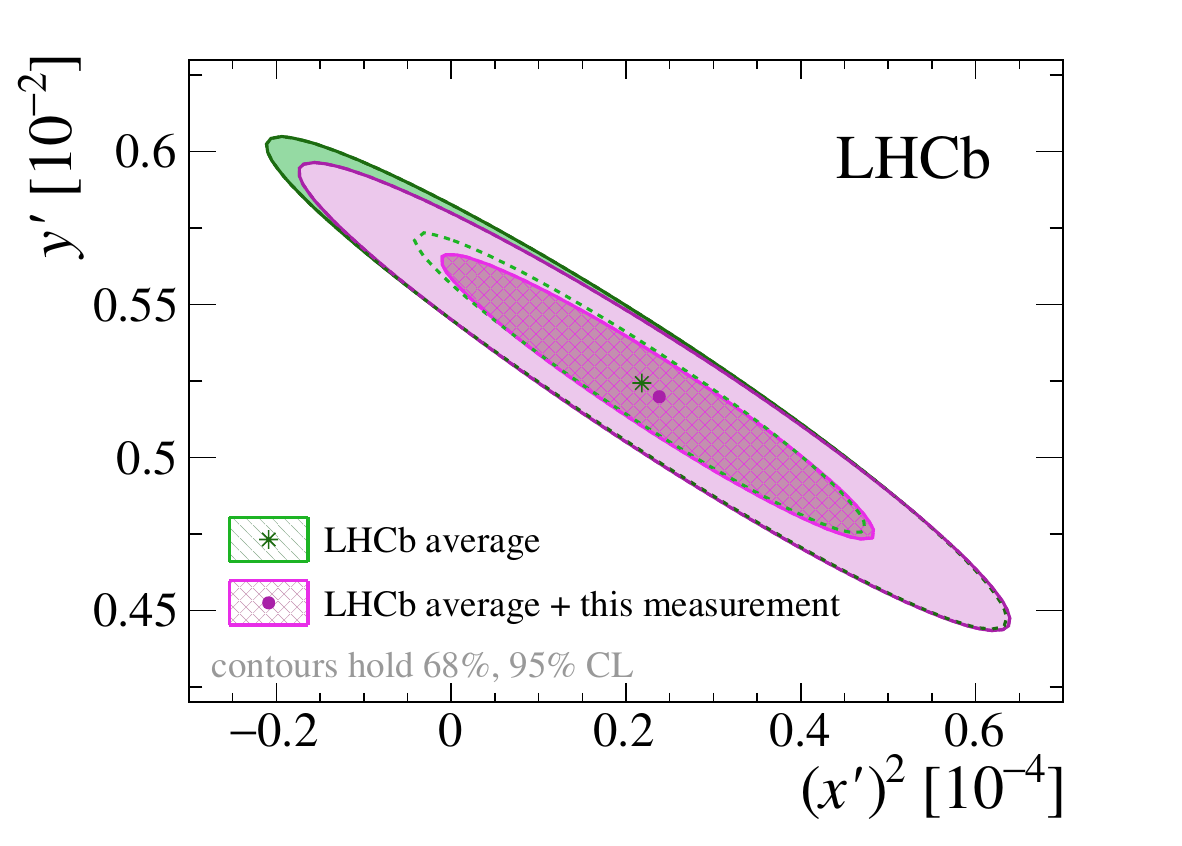}
    \caption{Correlation between the mixing parameters $y^\prime$ and $(x^\prime)^2$ illustrating (left) the individual Run~1 (blue) and Run~2 (orange) double-tagged results, and (right) the $\lhcb$ average before (green) and after (pink) including this result.}
    \label{fig:combination}
\end{figure}

The precision of each parameter has improved by around a factor of two with respect to the previous measurement due to the increased sample size.
This can be further improved by including the previous measurement~\cite{LHCb-PAPER-2016-033}, corresponding to the \lhcb Run~1 dataset.
The compatibility of the Run~1 and Run~2 analyses is shown in Fig.~\ref{fig:combination}~(left).
The Run~1 analysis fit strategy is almost identical to the one presented here, with very similar nuisance parameters.
As such, measurements from the previous analysis enable a simple extension of the $\chisq$ function of Eq.~\ref{eqn:chi2function} to include the extra data points and nuisance parameters.
Table~\ref{tab:R1+2result} presents the fit result obtained with the inclusion of Run~1 data, which further improves the precision of each fit parameter by around $10\,\%$.

Analysis of prompt \Dstarp decays produces complementary results to that of double-tagged decays.
Measurements with the two samples differ in strategy but complement each other when combined.
The selection of promptly produced \Dstarp decays leads to a suppression of low \Dz decay times, while the double-tagged sample has optimal efficiency to reconstruct the lowest decay times.
Overlap between the two Run~2 datasets is estimated by further applying, where possible, all the selection criteria detailed in Ref.~\cite{LHCb-PAPER-2024-008} to the double-tagged signal dataset.
This results in a maximum contamination in the sample containing selected double-tagged WS candidates of about $4\,\%$, which can be safely neglected in this or in future combinations.

The prompt data results are those reported, in an alternative parametrisation, in Table~3 of Ref.~\cite{LHCb-PAPER-2024-008}.
Appendix~\ref{sec:appA} translates the results obtained here to this parametrisation for comparison.
The addition of the results of this analysis reduces the parameter uncertainties obtained from promptly produced \Dstarp decays by ${4.8-6.4\,\%}$, as illustrated in Fig.~\ref{fig:combination}~(right).
The improvement is substantial considering the double-tagged yield is just $\sim1\,\%$ that of the prompt.
In addition to the complementary decay-time coverage, this is the result of different correlation coefficients and sample purities.

\begin{table}
    \centering
    \caption{Result including data collected during Run~1 and Run~2 of the \lhc. The first uncertainty is statistical and the second systematic. Correlations include both statistical and systematic contributions.}
    \begin{tabular}{ c R | S[table-format=3.1] S[table-format=3.1] S[table-format=3.1] S[table-format=3.1] S[table-format=3.1] S[table-format=3.1] }
        \hline
        \multirow{2}{*}{Parameter} & \multicolumn{1}{c}{\multirow{2}{*}{Value}} & \multicolumn{6}{c}{Correlations [\%]} \\
        \multicolumn{2}{c}{} & $\RDp$ & $\yp$ & $\xsqp$ & $\RDm$ & $\ym$ & ${(x^{\prime-})}^{2}$ \\
        \hline
        $\RDp$  & $( 350.0 \pm 6.9 \pm 2.3) \times 10^{-5}$ &  100.0 &  -74.9 &   62.4 &   -1.3 &   -0.0 &    0.0 \\
        $\yp$   & $(   4.1 \pm 2.0 \pm 0.3) \times 10^{-3}$ &        &  100.0 &  -94.3 &   -0.0 &   -0.0 &   -0.0 \\
        $\xsqp$ & $(   0.8 \pm 1.5 \pm 0.1) \times 10^{-4}$ &        &        &  100.0 &   -0.0 &   -0.0 &    0.0 \\
        $\RDm$  & $( 344.0 \pm 7.0 \pm 2.3) \times 10^{-5}$ &        &        &        &  100.0 &  -74.5 &   62.9 \\
        $\ym$   & $(   6.8 \pm 2.1 \pm 0.3) \times 10^{-3}$ &        &        &        &        &  100.0 &  -94.6 \\
        $\xsqm$ & $(  -0.5 \pm 1.7 \pm 0.1) \times 10^{-4}$ &        &        &        &        &        &  100.0 \\
        \hline
    \end{tabular}
    \label{tab:R1+2result}
\end{table}

\section{Conclusions}

The time-dependent ratio of the wrong- to right-sign ${\Dz\to\Kpm\pimp}$ decay rates is determined with data collected by the \lhcb experiment between 2016 and 2018, at a proton-proton centre-of-mass energy of $13\tev$, corresponding to an integrated luminosity of ${5.4\invfb}$.
The analogous decays of $\Dzb$ mesons are studied simultaneously to probe \CP violation in both the decay and flavour oscillations.
The flavour of the intermediate neutral meson at production is inferred from the charges of the two spectator particles in preceding ${\Bbar\to\theDstarp\mun X}$ and ${\theDstarp\to\Dz\pip}$ decays.
Uncertainties are predominantly comprised of statistical contributions with the most prominent systematic contribution relating to the modelling of the invariant-mass distribution from which decay rates are extracted.
The absolute precision of the mixing and \CP-violation parameters has improved by more than a factor of two with respect to the equivalent analysis performed on Run~1 data.
All results are consistent with \CP\ symmetry.
Combination with results obtained from promptly produced \Dstarp decays, which are compatible with the results presented here, leads to a precision that is improved significantly more than expected from the sample size alone.
This result completes the series of analyses of WS ${\Dz\to\kaon\pion}$ decays~\cite{LHCb-PAPER-2013-053,LHCb-PAPER-2016-033,LHCb-PAPER-2024-008} by LHCb with Run~1 and Run~2 data.

% Do not include this in any draft (just for information in the template)
% \input{acknowledgements_intro}
% Comment this in for paper drafts; do not include this in analysis note, conference and figure reports
\section*{Acknowledgements}
%
% These Acknowledgements valid from 3-May-2019
%
\noindent We express our gratitude to our colleagues in the CERN
accelerator departments for the excellent performance of the LHC. We
thank the technical and administrative staff at the LHCb
institutes.
We acknowledge support from CERN and from the national agencies:
CAPES, CNPq, FAPERJ and FINEP (Brazil); 
MOST and NSFC (China); 
CNRS/IN2P3 (France); 
BMBF, DFG and MPG (Germany); 
INFN (Italy); 
NWO (Netherlands); 
MNiSW and NCN (Poland); 
MCID/IFA (Romania); 
%MSHE (Russia); 
MICIU and AEI (Spain);
SNSF and SER (Switzerland); 
NASU (Ukraine); 
STFC (United Kingdom); 
DOE NP and NSF (USA).
We acknowledge the computing resources that are provided by CERN, IN2P3
(France), KIT and DESY (Germany), INFN (Italy), SURF (Netherlands),
PIC (Spain), GridPP (United Kingdom), 
%RRCKI and Yandex LLC (Russia), 
CSCS (Switzerland), IFIN-HH (Romania), CBPF (Brazil),
and Polish WLCG (Poland).
We are indebted to the communities behind the multiple open-source
software packages on which we depend.
Individual groups or members have received support from
ARC and ARDC (Australia);
Key Research Program of Frontier Sciences of CAS, CAS PIFI, CAS CCEPP, 
Fundamental Research Funds for the Central Universities, 
and Sci. \& Tech. Program of Guangzhou (China);
Minciencias (Colombia);
EPLANET, Marie Sk\l{}odowska-Curie Actions, ERC and NextGenerationEU (European Union);
A*MIDEX, ANR, IPhU and Labex P2IO, and R\'{e}gion Auvergne-Rh\^{o}ne-Alpes (France);
%RFBR, RSF and Yandex LLC (Russia);
AvH Foundation (Germany);
ICSC (Italy); 
%GVA, XuntaGal, GENCAT, Inditex, InTalent and Prog.~Atracci\'on Talento, CM (Spain);
Severo Ochoa and Mar\'ia de Maeztu Units of Excellence, GVA, XuntaGal, GENCAT, InTalent-Inditex and Prog. ~Atracci\'on Talento CM (Spain);
SRC (Sweden);
the Leverhulme Trust, the Royal Society
 and UKRI (United Kingdom).

\section*{Appendices}

\appendix

\section{Alternative parametrisation}
\label{sec:appA}

\begin{table}[t]
    \centering
    \caption{Result of the fully parametrised \chisq fit expressed in the alternative parametrisation. Uncertainties and correlations include both statistical and systematic contributions.}
    \begin{tabular}{ c M | S[table-format=3.1] S[table-format=3.1] S[table-format=3.1] S[table-format=3.1] S[table-format=3.1] S[table-format=3.1] }
        \hline
        \multirow{2}{*}{Parameter} & \multicolumn{1}{c}{\multirow{2}{*}{Value}} & \multicolumn{6}{c}{Correlations [\%]} \\
        \multicolumn{2}{c}{} & $\RD$ & $\ckp$ & $\ckpp$ & $\AD$ & $\Dckp$ & $\Dckpp$ \\
        \hline
        $\RD$    & $( 347.2 \pm 5.8) \times 10^{-5}$ &  100.0 &  -76.7 &   61.5 &   -1.5 &    1.6 &    0.1 \\
        $\ckp$   & $(   5.8 \pm 1.6) \times 10^{-3}$ &        &  100.0 &  -92.4 &    3.3 &   -5.0 &    2.8 \\
        $\ckpp$  & $(   0.9 \pm 2.6) \times 10^{-5}$ &        &        &  100.0 &   -1.3 &    2.8 &   -1.9 \\
        $\AD$    & $(   2.3 \pm 1.7) \times 10^{-2}$ &        &        &        &  100.0 &  -75.6 &   60.6 \\
        $\Dckp$  & $(  -2.3 \pm 1.6) \times 10^{-3}$ &        &        &        &        &  100.0 &  -92.4 \\
        $\Dckpp$ & $(   2.1 \pm 2.6) \times 10^{-5}$ &        &        &        &        &        &  100.0 \\
        \hline
    \end{tabular}
    \label{tab:alternative_paramertrisation}
\end{table}
\begin{table}[t]
    \centering
    \caption{Result including data collected during Run~1 and Run~2 of \lhcb in the alternative parametrisation. Uncertainties and correlations include both statistical and systematic contributions.}
    \begin{tabular}{ c M | S[table-format=3.1] S[table-format=3.1] S[table-format=3.1] S[table-format=3.1] S[table-format=3.1] S[table-format=3.1] }
        \hline
        \multirow{2}{*}{Parameter} & \multicolumn{1}{c}{\multirow{2}{*}{Value}} & \multicolumn{6}{c}{Correlations [\%]} \\
        \multicolumn{2}{c}{} & $\RD$ & $\ckp$ & $\ckpp$ & $\AD$ & $\Dckp$ & $\Dckpp$ \\
        \hline
        $\RD$    & $( 347.0 \pm 5.1) \times 10^{-5}$ &  100.0 &  -75.2 &   60.4 &   -1.4 &    1.3 &   -0.5 \\
        $\ckp$   & $(   5.5 \pm 1.5) \times 10^{-3}$ &        &  100.0 &  -92.5 &    1.9 &   -3.5 &    2.4 \\
        $\ckpp$  & $(   1.2 \pm 2.4) \times 10^{-5}$ &        &        &  100.0 &   -1.0 &    2.4 &   -2.4 \\
        $\AD$    & $(   0.9 \pm 1.5) \times 10^{-2}$ &        &        &        &  100.0 &  -74.1 &   59.7 \\
        $\Dckp$  & $(  -1.3 \pm 1.5) \times 10^{-3}$ &        &        &        &        &  100.0 &  -92.5 \\
        $\Dckpp$ & $(   1.2 \pm 2.4) \times 10^{-5}$ &        &        &        &        &        &  100.0 \\
        \hline
    \end{tabular}
    \label{tab:alternative_paramertrisation combination}
\end{table}

The analysis of Run~2 prompt $\Dz\to\Kpm\pimp$ decays measures ratios defined by
\begin{align}
    \frac{\left|\langle{\Kp\pim}|\mathbf{H}|{\Dz(t)}\rangle\right|^2}{\left|\langle{\Kp\pim}|\mathbf{H}|{\Dzb(t)}\rangle\right|^2} &= \bar{R}^{+}(t), \\
    \frac{\left|\langle{\Km\pip}|\mathbf{H}|{\Dzb(t)}\rangle\right|^2}{\left|\langle{\Km\pip}|\mathbf{H}|{\Dz(t)}\rangle\right|^2} &= \bar{R}^{-}(t),
\end{align}
alternative to those defined in Eqs.~\ref{eqn: R+}~and~\ref{eqn: R-}. This leads to an alternative parametrisation of the mixing and \CP violation parameters, given by
\begin{equation}
    \bar{R}^{\pm}(t) \approx \RD(1 \pm \AD) + \sqrt{\RD(1 \pm \AD)}(\ckp \pm \Dckp)\left(\frac{t}{\tau_{\Dz}}\right) + (\ckpp \pm \Dckpp)\left(\frac{t}{\tau_{\Dz}}\right)^{2},
\end{equation}
where
\begin{align}
    \RD &= \frac{\RDp+\RDm}{2}, \\
    \ckp &= \frac{\yp+\ym}{2}, \\
    \ckpp &= \frac{1}{2}\left[\frac{\xsqp + (\yp)^{2}}{4} + \frac{\xsqm + (\ym)^{2}}{4}\right], \\
    \AD &= \frac{\RDp-\RDm}{\RDp+\RDm}, \\
    \Dckp &= \frac{\yp-\ym}{2}, \\
    \Dckpp &= \frac{1}{2}\left[\frac{\xsqp + (\yp)^{2}}{4} - \frac{\xsqm + (\ym)^{2}}{4}\right].
\end{align}
To facilitate future combinations, the results obtained in Tables~\ref{tab:result_ACPVA}~and~\ref{tab:R1+2result} are presented in the alternative parametrisation in Tables~\ref{tab:alternative_paramertrisation}~and~\ref{tab:alternative_paramertrisation combination}.

% This should be taken out in the final paper
% \input{supplementary-app}

\clearpage
\addcontentsline{toc}{section}{References}
%\setboolean{inbibliography}{true}
\bibliographystyle{LHCb}
\bibliography{main,standard,LHCb-PAPER,LHCb-CONF,LHCb-DP,LHCb-TDR}

\newpage
% LHCb collaboration author list
% Data extracted on September 25th, 2024 at 3:06pm for paper reference LHCb-PAPER-2024-044
\centerline
{\large\bf LHCb collaboration}
\begin
{flushleft}
\small
R.~Aaij$^{38}$\lhcborcid{0000-0003-0533-1952},
A.S.W.~Abdelmotteleb$^{57}$\lhcborcid{0000-0001-7905-0542},
C.~Abellan~Beteta$^{51}$,
F.~Abudin{\'e}n$^{57}$\lhcborcid{0000-0002-6737-3528},
T.~Ackernley$^{61}$\lhcborcid{0000-0002-5951-3498},
A. A. ~Adefisoye$^{69}$\lhcborcid{0000-0003-2448-1550},
B.~Adeva$^{47}$\lhcborcid{0000-0001-9756-3712},
M.~Adinolfi$^{55}$\lhcborcid{0000-0002-1326-1264},
P.~Adlarson$^{82}$\lhcborcid{0000-0001-6280-3851},
C.~Agapopoulou$^{14}$\lhcborcid{0000-0002-2368-0147}
% ...

{\footnotesize \it

% ...
$^{14}$Universit{\'e} Paris-Saclay, CNRS/IN2P3, IJCLab, Orsay, France\\
% ...
$^{38}$Nikhef National Institute for Subatomic Physics, Amsterdam, Netherlands\\
% ...
$^{47}$Instituto Galego de F{\'\i}sica de Altas Enerx{\'\i}as (IGFAE), Universidade de Santiago de Compostela, Santiago de Compostela, Spain\\
% ...
$^{51}$Physik-Institut, Universit{\"a}t Z{\"u}rich, Z{\"u}rich, Switzerland\\
% ...
$^{55}$H.H. Wills Physics Laboratory, University of Bristol, Bristol, United Kingdom\\
% ...
$^{57}$Department of Physics, University of Warwick, Coventry, United Kingdom\\
% ...
$^{60}$School of Physics and Astronomy, University of Glasgow, Glasgow, United Kingdom\\
$^{61}$Oliver Lodge Laboratory, University of Liverpool, Liverpool, United Kingdom\\
% ...
$^{69}$Syracuse University, Syracuse, NY, United States\\
% ...
$^{82}$Department of Physics and Astronomy, Uppsala University, Uppsala, Sweden, associated to $^{60}$\\
% ...
}
\end{flushleft}

\end{document}